\documentclass[twocolumn]{aastex631}

\hypersetup{linkcolor=blue,citecolor=blue,filecolor=cyan,urlcolor=magenta}
\usepackage{appendix,natbib}
\usepackage{amsmath}
\usepackage{courier}
\usepackage{booktabs}
\usepackage{upgreek}
\usepackage{rotating}
\usepackage{xcolor}

\newcommand\U{$u$}
\newcommand\G{$g$}
\newcommand\R{$r$}
\newcommand\I{$i$}
\newcommand\Z{$z$}
\newcommand\Y{$Y$}
\newcommand\Ks{$K_s$}
\newcommand\um{$\mu$m}
\newcommand{\dotdeg}{\rlap{.}^\circ}

\newcommand{\dotarcsec}{\rlap{.}''}
\newcommand\farmer{\texttt{The Farmer}}

\shorttitle{SHELA catalog}
\shortauthors{Leung et al.}

\begin{document}

\title{The {\it Spitzer}-HETDEX Exploratory Large Area Survey. IV. Model-Based Multi-wavelength Photometric Catalog}

\author[0000-0002-9393-6507]{Gene C. K. Leung}
\affiliation{Department of Astronomy, The University of Texas at Austin, 2515 Speedway, Stop C1400, Austin, TX 78712, USA}

\author[0000-0001-8519-1130]{Steven L. Finkelstein}
\affiliation{Department of Astronomy, The University of Texas at Austin, 2515 Speedway, Stop C1400, Austin, TX 78712, USA}

\author[0000-0003-1614-196X]{John~R.~Weaver}
\affil{Department of Astronomy, University of Massachusetts, Amherst, MA 01003, USA}

\author[0000-0001-7503-8482]{Casey Papovich}
\affiliation{Department of Physics and Astronomy, Texas A\&M University, College Station, TX, USA}
\affiliation{George P. and Cynthia Woods Mitchell Institute for Fundamental Physics and Astronomy, Texas A\&M University, College Station, TX, USA}

\author[0000-0003-2366-8858]{Rebecca L. Larson}
\affiliation{Department of Astronomy, The University of Texas at Austin, 2515 Speedway, Stop C1400, Austin, TX 78712, USA}

\author[0000-0003-4922-0613]{Katherine Chworowsky}
\affiliation{Department of Astronomy, The University of Texas at Austin, 2515 Speedway, Stop C1400, Austin, TX 78712, USA}

\author[0000-0002-1328-0211]{Robin Ciardullo}
\affiliation{Department of Astronomy and Astrophysics, The Pennsylvania State University, University Park, PA 16802, USA}
\affiliation{Institute for Gravitation and the Cosmos, The Pennsylvania State University, University Park, PA 16802, USA}

\author[0000-0003-1530-8713]{Eric Gawiser}
\affiliation{Department of Physics and Astronomy, Rutgers, The State University of New Jersey, 136 Frelinghuysen Rd., Piscataway, NJ 08854, USA}

\author[0000-0001-6842-2371]{Caryl Gronwall}
\affiliation{Department of Astronomy and Astrophysics, The Pennsylvania State University, University Park, PA 16802, USA}
\affiliation{Institute for Gravitation and the Cosmos, The Pennsylvania State University, University Park, PA 16802, USA}

\author[0000-0002-1590-0568]{Shardha Jogee}
\affiliation{Department of Astronomy, The University of Texas at Austin, 2515 Speedway, Stop C1400, Austin, TX 78712, USA}

\author[0000-0003-4032-2445]{Lalitwadee Kawinwanichakij}
\affiliation{Kavli Institute for the Physics and Mathematics of the Universe, The University of Tokyo, Kashiwa 277-8583, Japan (Kavli IPMU, WPI)}

\author[0000-0002-6748-6821]{Rachel S. Somerville}
\affiliation{Center for Computational Astrophysics, Flatiron Institute, 162 5th Avenue, New York, NY 10010, USA}

\author[0000-0002-0784-1852]{Isak G. B. Wold}
\affiliation{Astrophysics Science Division, NASA Goddard Space Flight Center, 8800 Greenbelt Rd, Greenbelt, MD 20771, USA}

\author[0000-0003-3466-035X]{L. Y. Aaron Yung}
\affiliation{Astrophysics Science Division, NASA Goddard Space Flight Center, 8800 Greenbelt Rd, Greenbelt, MD 20771, USA}


\begin{abstract}
We present a 0.3--4.5 \um\ 16-band photometric catalog for the {\it Spitzer}/HETDEX Exploratory Large-Area (SHELA) survey. SHELA covers a $\sim 27$ deg$^2$ field within the footprint of the Hobby-Eberly Telescope Dark Energy Experiment (HETDEX). Here we present new DECam imaging and a $rizK_s$-band-selected catalog of four million sources extracted using a fully model-based approach. We validate our photometry by comparing with the model-based DECam Legacy Survey. We analyze the differences between model-based and aperture photometry by comparing with the previous SHELA catalog, finding that our model-based photometry can measure point sources to fainter fluxes and better capture the full emission of resolved sources. The catalog is $80\%$ ($50\%$) complete at $riz \sim 24.7$ ($25.1$) AB mag, and the optical photometry reaches a $5\sigma$ depth of $\sim 25.5$ AB mag. We measure photometric redshifts and achieve $1\sigma$ scatter of $\Delta z/(1+z)$ of 0.04 with available spectroscopic redshifts at $0 \le z \le 1$. This large area, multi-wavelength photometric catalog, combined with spectroscopic information from HETDEX, will enable a wide range of extragalactic science investigations.

\end{abstract}


\section{Introduction}\label{sec:intro}

Hobby-Eberly Telescope Dark Energy Experiment (HETDEX) \citep{hetdex} is an unbiased integral-field spectroscopic survey aiming to measure Ly$\alpha$ emission from one million galaxies at $z=1.9-3.5$ in a 540 deg$^2$ region. The SHELA survey \citep{pap16} is a wide-field multi-wavelength photometric survey targeting a $\sim 27$ deg$^2$ field within the footprint of HETDEX\null. It is designed to study the physical processes, intrinsic and environmental, driving the growth of galaxies from $z=1.9-3.5$. The SHELA survey has obtained optical imaging using DECam, near-infrared (NIR) imaging using NEWFIRM, and mid-infrared imaging using {\it Spitzer}/IRAC, and is supplemented by a large amount of publicly available ground-based imaging data. 

Early SHELA observations have resulted in three published catalogs. \citet{pap16} presented a 3.6 and 4.5 \um ~catalog of the SHELA {\it Spitzer}/IRAC dataset. \citet{wol19} used data from the first phase of SHELA DECam observations and presented an $riz$-selected catalog with DECam $ugriz$ photometry along with the existing IRAC data and $JK_s$ data from the VICS82 survey \citep{vics82}. \citet{ste21} presented a $K_s$-selected catalog combining the previous dataset with $K_s$-band observations from the NEWFIRM HETDEX Survey. Since then, we have completed the second and final phase of our DECam imaging campaign in SHELA. In this paper, we present a new photometric catalog by combining our complete 27-deg$^2$ DECam imaging dataset with the existing SHELA NEWFIRM and {\it Spitzer}/IRAC observations. Apart from the new DECam imaging data that increase both the width and depth of our survey, we discuss below several key differences of this new catalog from the previous ones. 

In this catalog, we benefit from a number of publicly available imaging datasets in addition to our own imaging campaign. A crucial addition from the previous catalogs is the Hyper Suprime-Cam Subaru Strategic Program \citep[HSC-SSP,][]{hscpdr3}, which has observed the SHELA field in \G\R\I\Z$y$ ~as a part of its autumn field. The publicly available HSC-SSP images in the SHELA field are $\sim 0.5-1$ mag deeper than the DECam images in \citet{wol19}. Combining these deep HSC-SSP images with our DECam dataset allows precise measurement of the optical spectral energy distribution (SED) and constrains the physical properties of galaxies. For example, deeper \G -band imaging is important in reducing the contamination rate in the selection of $z \gtrsim 3-5$ galaxies.

The full coverage of the SHELA field with both DECam and NEWFIRM supplemented by HSC-SSP now allows us to select galaxies using the optical and NIR bands simultaneously. As opposed to $riz$ only in \citet{wol19}, we will be able to detect, using the $K_s$ band, very red galaxies that peak in the NIR, such as massive quiescent galaxies, dusty star-forming galaxies and $z \ge 7$ Lyman break galaxies. On the other hand, the $K_s$-band selected catalog in \citet{ste21} is limited by its depth of $\sim 22.4$ mag, and the inclusion of deep optical imaging in this catalog that is $\sim 2-3$ mag deeper will significantly increase the completeness, such as for $z \sim 2 - 5$ star-forming galaxies whose rest-frame UV emission falls in the optical wavelengths.

The previous SHELA catalogs were created entirely or partially based on aperture photometry using tools such as \texttt{SourceExtractor} \citep{ber96}. In such an approach, the images in each band will be homogenized to a common PSF, leading to the degradation of spatial resolution in the images. An alternative approach is to model the light profile of sources convolved with the measured PSF. Recently, the image modeling code \texttt{the Tractor} \citep{tractor} has been developed to to perform such model-based photometry. It has since been implemented by a number of imaging surveys, including the DECam Legacy Survey \citep[DECaLS,][]{decals} and COSMOS2020 catalog \citep{wea22}. The model-based approach is crucial in extracting IRAC photometry, which has a substantially larger PSF than the optical bands. For this catalog, we extract fully model-based photometry using the software package \farmer ~\citep{wea22}, which generates photometry using \texttt{the Tractor}.

The large area, multi-wavelength photometric catalog of SHELA will enable a variety of extragalactic studies. In particular, studies that focus on galaxies spanning a wide range of physical parameters, galaxies residing in a variety of environments or the search for rare objects will benefit the most from the large area of SHELA.

This paper is organized as follows. In Section \ref{sec:obs}, we describe our observations and the composition of our imaging dataset. Section \ref{sec:data} describes the data reduction procedures for the images. The photometry extraction using \farmer ~and the photometry validation process are described in Section \ref{sec:phot}. Section \ref{sec:photz} presents the photometric redshift measurements. We give a description of the catalog published with this paper in Section \ref{sec:cat}. In this paper we assume a \citet{pla20} cosmology of $H_0=67.4 ~ \mathrm{km~s}^{-1} ~\mathrm{Mpc}^{-1}$, $\Omega_\mathrm{m}=0.315$ and $\Omega_\mathrm{\Lambda} = 0.685$.

\begin{figure*}[!htbp]
	\centering
		\includegraphics[width=\textwidth]{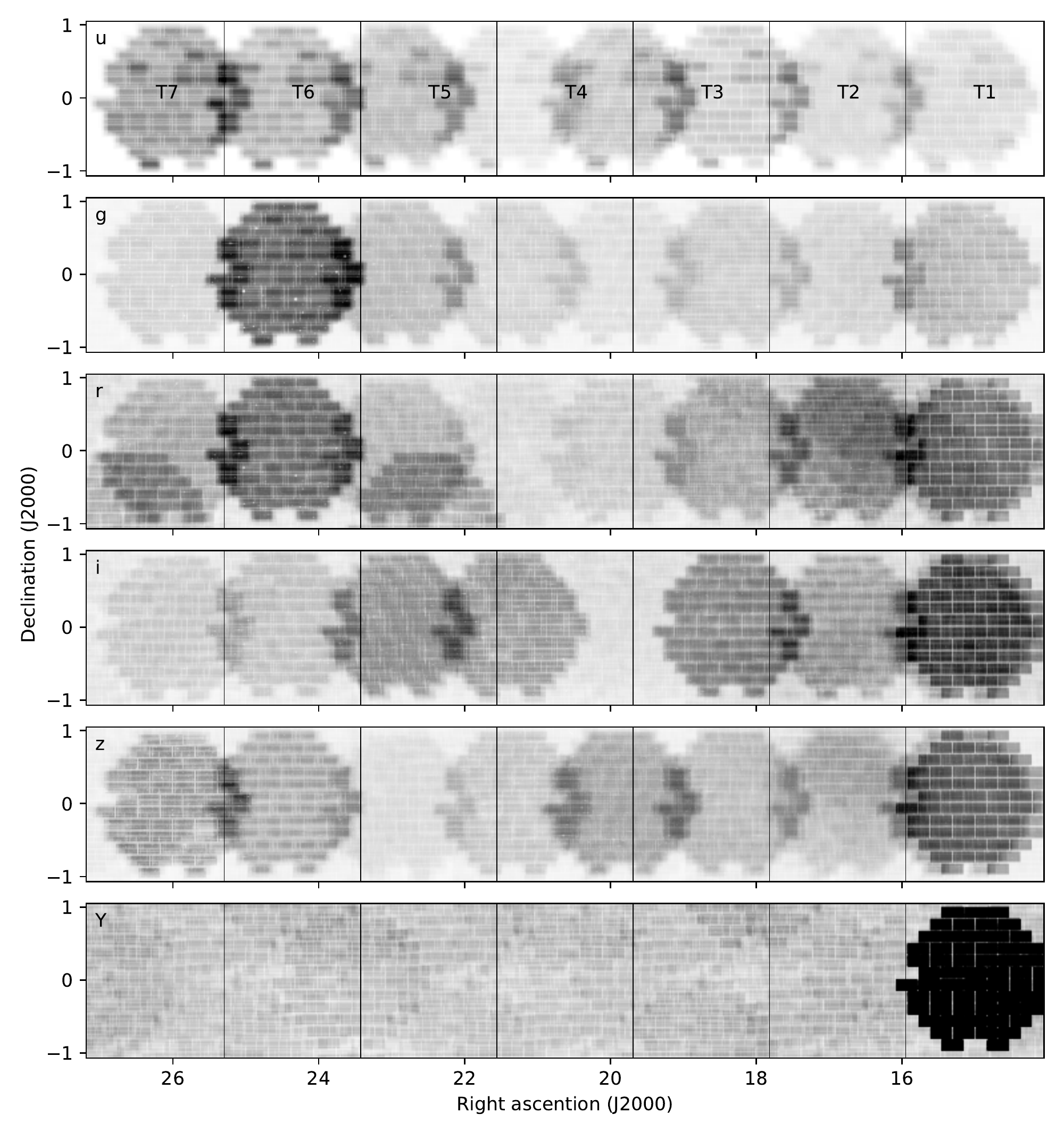}
		\caption{SHELA DECam $ugrizY$ weight maps demonstrating the survey's footprint and relative depths. The field is covered by eight DECam pointings, and is divided into seven tiles of $1\dotdeg98 \times 2\dotdeg145$, labelled T1 to T7, so that the tiles are roughly square. Our survey combines $ugriz$ images from our observations with $grizY$ archival images released prior to March 3 2021. We note that T1 has substantially deeper $Y$-band imaging due to a large number of repeated exposures in the archive.}
		\label{fig:footprint}
\end{figure*}

\section{Observations and Data} \label{sec:obs}
We observed the SHELA field using the DECam imager on the Blanco 4m telescope at CTIO in the \U-, \G-, \R-, \I-, \Z- filters. New data were obtained over 3 nights in the 2019B semester (PI: Papovich; NOAO PID: 2019B-0080). Our program also includes data obtained over 5 nights in the 2013B semester presented in \citet{wol19}. The complete program consists of eight slightly overlapping pointings covering a total area of 23 deg$^2$. The 2013B data covered six of the eight pointings, while the new 2019B data completed the last two pointings and obtained additional exposures in the previous positions. 

The SHELA field has been observed with DECam by other programs, most notably the Dark Energy Survey (DES, \citealt{des}) and DECaLS \citep{decals}.
We supplement our data with archival DECam imaging data within the footprint of the SHELA field in the \U\G\R\I\Z ~bands. We also include archival data in \Y ~band, which was not observed by our proprietary program. All such archival data available through the NOIRLab Astro Data Archive\footnote{https://astroarchive.noirlab.edu/} released prior to March 3 2021 were included for this catalog. This results in a total of [200, 719, 735, 688, 720, 468] exposures in the [\U, \G, \R, \I, \Z, \Y] bands. We show the DECam $ugrizY $ weight maps in the SHELA field in Figure \ref{fig:footprint} to demonstrate the footprint and relative depths of the DECam data.

\begin{figure*}[!t]
	\centering
		\includegraphics[width=\textwidth]{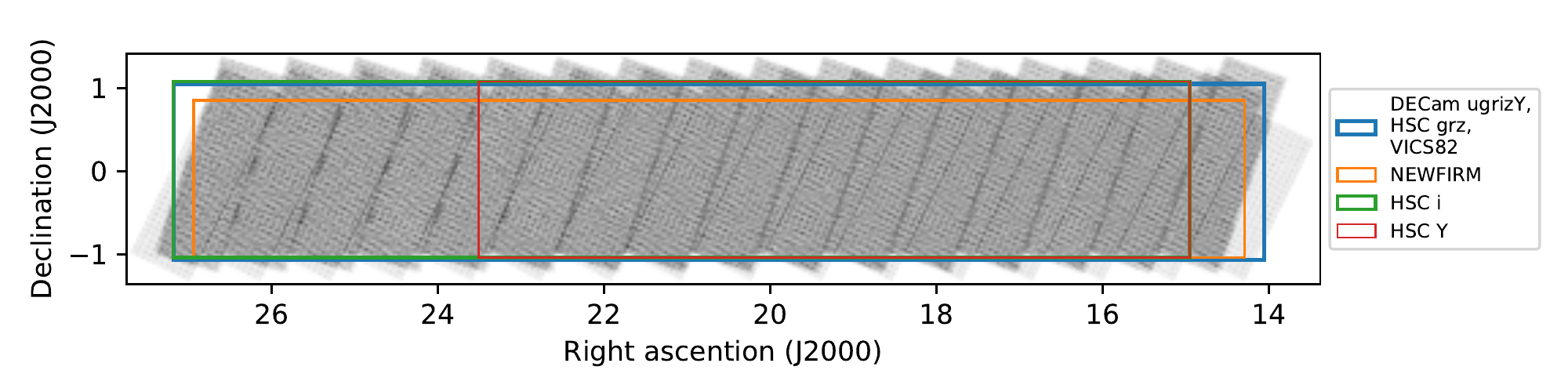}
		\caption{Exposure map of the IRAC imaging overlaid with the coverage of data from all the surveys used in this catalog. The area included this catalog is defined by the coverage of our own DECam program, which is completely filled by the HSC $grz$-bands and VICS82 imaging.}
		\label{fig:surveys}
\end{figure*}

\begin{figure*}[!t]
	\centering
		\includegraphics[width=\textwidth]{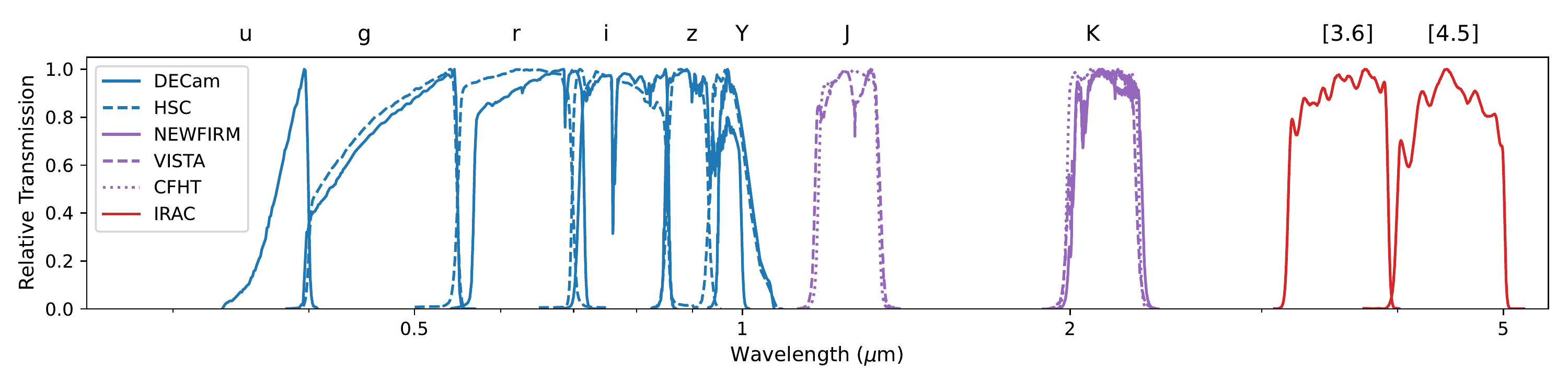}
		\caption{Ralative transmission curves for the photometric bands used. The effects of the filter, atmosphere, detector and optics are included.}
		\label{fig:filters}
\end{figure*}

We incorporate multi-wavelength imaging data from surveys overlapping with the SHELA field. Previous SHELA programs have obtained data in the near- and mid-infrared wavelengths. This includes NEWFIRM \Ks -band images from the NEWFIRM HETDEX Survey \citep{ste21} and mid-infrared mosaics in 3.6 \um\ and 4.5 \um\ with {\it Spitzer}/IRAC presented in \citet{pap16}. We also make use of publicly available data from PDR3 of HSC-SSP \citep{hscpdr3} within the SHELA field. Finally we include $J$ and (shallower) \Ks\ images from the VICS82 survey \citep{vics82} obtained with VISTA and CFHT\null. We show the coverage of all the surveys included in this catalog in Figure \ref{fig:surveys}. Figure \ref{fig:filters} shows the filter transmission curves of all the photometric bands used in this catalog.


\section{Data Reduction} \label{sec:data}
To perform model-based photometry with \farmer, images in each photometric band need to be resampled and stacked to the same pixel grids. In this section, we describe the procedures to obtain the stacked images and zero-points in each photometric band.

\subsection{DECam Images}
We begin the stacking of DECam images with the NOAO DECam Community Pipeline (CP) resampled images of each individual exposure. The NOAO DECam CP resampled images have been photometrically and astrometrically calibrated to a common pixel scale of $0\farcs27$ per pixel. The DECam exposures in our data set were observed through a seven-year period, and the images were reduced using different versions of the CP\null. Therefore, we re-calibrate the astrometry and flux scaling of each image uniformly prior to stacking.

\subsubsection{Astrometric and Flux-scaling Calibration}
\label{sec:cal}
We tie the astrometry of each pre-stack image to the Gaia EDR3 catalog \citep{gai21}. For each image, we first generate an initial source catalog using \texttt{SEP} \citep{bar16}, a Python library that implements the core algorithm of \texttt{SourceExtractor} \citep{ber96}. We match sources in this catalog to the reported coordinates of astrometry stars in the Gaia EDR3 catalog. We first identify stars using their Gaia catalog colors \citep{bai19}. We then select astrometry stars as those with a {\it G} band S/N greater than 10 and proper motion less than $10'' \mathrm{yr}^{-1}$ in the Gaia EDR3 catalog, and no extraction flags in our \texttt{SEP} catalog. We then calculate the median $x$- and $y$-offset required to match the good stars to the Gaia coordinates in each image. 

Before stacking, it is common to first scale the images to a common count level to account for differences in exposure time and extinction. To do so, we measure a flux scaling factor for each image from a set of PSF stars. PSF stars are defined as astrometry stars from above with no neighbours within $5''$ in our initial catalog. We measure the flux of these PSF stars in a $2\dotarcsec5$ radius aperture, as well as an aperture that, on median, contains $70\%$ of the $2\dotarcsec5$ aperture flux. The $2\dotarcsec5$ aperture is a proxy for the total flux except for very bright stars, but it includes substantial noise for fainter stars. The $70\%$ aperture provides a more robust measurement of flux for stars of a range of brightness. We divide the $70\%$ aperture flux by a factor of 0.7 to obtain the total flux measurements of the PSF stars. 

We then calculate the flux scaling factor required to scale each image to the count level of a reference image. For each band, the reference image is selected in the T4 tile at the center of the SHELA field, and is required to have seeing in the lower half of all images and at least 1000 PSF stars identified. The count level of the reference image is then propagated to the rest of the SHELA field from a list of overlapping images satisfying the seeing and PSF star criteria, resulting in a catalog of PSF stars at the reference count level for the entire SHELA field. The flux scaling factor is calculated by the median ratio between the catalog counts and measured counts of the PSF stars in the image.

\subsubsection{Image Stacking} \label{sec:stack}

We stack the exposures to seven tiles of $1\dotdeg98 \times 2\dotdeg145$ with a pixel scale of $0\dotarcsec27$ per pixel using \texttt{SWarp} \citep{ber10}. Each tile overlaps with its neighbouring tile by approximately $3'$. We apply the astrometric correction and flux scaling factors obtained above using an external header file. We remove any CCDs within a DECam exposure if the number of bad pixels in the data quality mask (DQM) exceeds $20\%$ of the total number of pixels. We also exclude the S7 CCD from all exposures as it is known to have a defective amplifier. This leaves us with 61 working CCDs for most DECam pointings. For a small number of exposures, some other CCDs display a strong variable background and discontinuity between the two amplifiers, and are removed from our stacking. Bleed trails from bright stars that are not removed by the CP are identified and masked. We then stack the exposures using the surface-brightness-optimized weights following \citet{gaw06}:
\begin{equation}
    w_i^\mathrm{SB} = \left( \frac{1}{p_i \mathrm{rms}^i} \right)^2,
\end{equation}
where $p_i$ is the flux scaling factor and rms$_i$ is the pixel-by-pixel rms of the background in the science image. 
Pixels with a nonzero DQM value are assigned a weight of zero. We utilize the recommended \texttt{Lanczos3} interpolation function to resample and co-add the images.

\subsection{Non-DECam Images}

We construct non-DECam images in the same tiles and tangent points. The source images in these bands have been previously calibrated, and stacked to different tile schemes than our SHELA tiles. We resample and moasic the source images to match with the DECam stacks. We first update the astrometry of each image to match Gaia EDR3 using the same astrometric correction procedures described in Section \ref{sec:cal}. We do not calculate a flux-scaling factor or surface-brightness optimized weight, as these are calibrated stacked images. We use the available DQM and weight maps for masking and weighting. The exceptions are NHS \Ks ~and IRAC images, for which only exposure maps are available instead of weight maps. For these bands, we scale the exposure maps with the background rms to create inverse-variance weight maps. The publicly available HSC-SSP images are previously stacked to ``patches'' with overlapping edges with neighbouring ones, so we trim each image accordingly to avoid double-counting weights in those regions. The images are then resampled and mosaiced using the same \texttt{SWarp} configurations as the DECam images.

\subsection{Detection Images}\label{sec:det_im}

For each tile, we construct a detection image by co-adding the DECam \R, \I, \Z ~and NHS \Ks ~bands using the \texttt{CHI-MEAN} co-addition in \texttt{SWarp}. The \texttt{CHI-MEAN} co-addition method is the normalized quadrature sum of the signal-to-noise in each band offset to be centered at zero (see \citealt{drl18}). We select the \R, \I, \Z ~and \Ks ~bands to optimize the detection of higher redshift galaxies, as many will drop out in the \U ~and \G ~bands. 

\subsection{Zero-point Determination}

We calibrate the zero-point of the DECam images to the Pan-STARRS1 (PS1) DR2 catalog for the $grizY$~bands, and to the SDSS DR16 for the \U ~band, which is not covered by PS1. We first identify F0 stars with S/N $>10$ in the reference catalogs using a color cut on their catalog magnitudes. We select F0 stars because of their relatively flat SED and large number density. We determine the expected colors of an F0 star using the PS1 and SDSS filter transmission curves and the \citet{kur93} F0 SED template. We adopt the following color cuts for the PS1 and SDSS catalogs to obtain $>2000$ F0 stars in the SHELA field:
\begin{align}
    & (g_\mathrm{PS1} - r_\mathrm{PS1} - 0.11)^2 + (r_\mathrm{PS1} - i_\mathrm{PS1} + 0.04)^2 \nonumber \\ 
    + & (i_\mathrm{PS1} - z_\mathrm{PS1} + 0.08)^2 + (z_\mathrm{PS1} - y_\mathrm{PS1} + 0.04)^2 \nonumber \\
    < & 0.27^2
\end{align}
\begin{align}
    & (u_\mathrm{SDSS} - g_\mathrm{SDSS} - 0.96)^2 + (g_\mathrm{SDSS} - r_\mathrm{SDSS} - 0.14)^2 \nonumber \\
    + & (r_\mathrm{SDSS} - i_\mathrm{SDSS} + 0.03)^2 + (i_\mathrm{SDSS} - z_\mathrm{SDSS} + 0.09)^2 \nonumber \\
    < & 0.2^2.
\end{align}

We calculate the zero-points of the \G, \R, \I, \Z ~and \Y ~bands using the PS1 catalog and that of the \U ~band using the SDSS catalog following the procedures in \citet{wol19}. Since the DECam filter transmission curves are not identical to those of PS1 or SDSS, we calculate the expected DECam magnitude using a linear color relation with the reference catalog magnitude. For example, in the \G ~band, we assume
\begin{equation}
    g^{AB}_\mathrm{PS1} = g_\mathrm{DECam} + \mathrm{ZPT} + \alpha (g^\mathrm{AB}_\mathrm{PS1} - r^\mathrm{AB}_\mathrm{PS1}),
\end{equation}
where $g_\mathrm{DECam}$ is the \G ~band instrumental DECam magnitude, ZPT is the zero-point, and $\alpha$ is the color slope. We solve for the zero-point and the color slope in each band and tile by performing a least square fit to the linear relation. The DECam magnitude is measured using the $70\%$ flux aperture described in Section \ref{sec:cal}. 

For the \U ~band, a sharp 4000\AA ~break in stellar spectra makes the color a poor predictor of the DECam flux \cite[see][]{wol19}. We instead compute the zero point by computing the expected magnitude offset between the DECam and SDSS filters ($\Delta u_\mathrm{filter}$) from the filter transmission curves and F0 SED template. The zero-point is given by
\begin{equation}
    \mathrm{ZPT} = \mathrm{median}(u_\mathrm{SDSS}^{AB} - u_\mathrm{DECam} - \Delta u_\mathrm{filter}).
\end{equation}
We subtract 0.04 from the zero-point to bring the SDSS magnitudes in alignment with AB magnitudes\footnote{\url{https://www.sdss.org/dr16/algorithms/fluxcal/}}.

The HSC-SSP, NHS, VICS82 and IRAC stacks are constructed from calibrated images, and their zero-points are preserved throughout the stacking procedure. We verify the final zero-points using a similar procedure as above, and find that they are consistent within uncertainty. We thus adopt the documented zero-points for these bands.


\section{Source Detection and Photometry} \label{sec:phot}

\subsection{\farmer}

Source detection and photometry are performed using \farmer , a software package that utilizes \texttt{The Tractor} to perform source modeling and photometry on multi-wavelength data. A detailed description of \farmer ~software package can be found in \citet{wea22}. To facilitate parallel computation, each tile is divided into 156 ``bricks'' with dimensions of approximately $10' \times 10'$. Source detection, modeling and forced photometry are performed in each brick independently. In this section, we describe each step in \farmer .

\subsubsection{Source Detection}

Before source detection, we mask extremely extended and resolved sources in the brick images since they can result in non-convergence in the modeling process and lead to inaccurate photometry of sources in their vicinity. We perform an initial source detection procedure on the detection image in each brick using \texttt{SEP} with a minimum detection area of 8000 pixels to create a segmentation map of very extended sources. The object regions in the segmentation map are dilated on each side by 15 pixels to create an extended object mask. In the HSC-SSP images, the extended object mask is dilated by 95 pixels because of the larger bright star halos for the instrument. The extended object masks are combined with the masks for the detection image and modeling images of each band.

Source detection in \farmer ~utilizes \texttt{SEP}. We use the weight maps generated in Section \ref{sec:stack} and masks generated above. The values of the source detection parameters used are shown in Table \ref{tab:sep}. After source detection, \farmer ~uses the segmentation map to identify crowded regions with multiple nearby sources, and groups them into ``blobs''. The blobs are the smallest units for source modeling, meaning that all the sources in a blob are modelled simultaneously.

\begin{deluxetable}{cc} \label{tab:sep}
\tablecaption{Source Detection Parameters}
\tablehead{
\colhead{Parameter} & \colhead{Value}
}
\startdata
\texttt{THRESH} & 1.5 \\
\texttt{MINAREA} & 3 \\
\texttt{DEBLEND\_NTHRESH} & 32 \\
\texttt{DEBLEND\_CONT} & 0.0001\\
\texttt{FILTER\_KERNEL} & gauss\_1.5\_3$\times$3.conv \\
\texttt{FILTER\_TYPE} & matched \\
\texttt{BW} & 128 \\
\texttt{BH} & 128 \\
\texttt{FW} & 3 \\
\texttt{FH} & 3 
\enddata
\end{deluxetable}

\subsubsection{PSF Creation}

To model the source profiles across multiple bands with inhomogeneous PSFs, we need to first model the PSF in each band and in each tile. This is achieved using \texttt{PSFEx}, which is integrated into \farmer . A catalog of bright sources is first constructed using \texttt{SourceExtractor} in each tile. Unsaturated point sources are then selected visually using the half-light radius versus apparent magnitude diagram. The selected point sources are then used for PSF modeling, where we allow spatial variations in each tile using a third degree polynomial. This allows a spatially dependent PSF model which accounts for variations in the image quality across each the tile.

\subsubsection{Source Modeling}

\begin{deluxetable}{cc}
\tablecaption{Farmer Decision Tree Parameters}\label{tab:mod}
\tablehead{
\colhead{Parameter} & \colhead{Value}
}
\startdata
\texttt{PS\_SG\_THRESH1} & 0.8 \\
\texttt{EXP\_DEV\_THRESH} & 0.5 \\
\texttt{CHISQ\_FORCE\_EXP\_DEV} & 1.05 \\
\texttt{CHISQ\_FORCE\_COMP} & 1.5
\enddata
\end{deluxetable}

\begin{figure}[!tbp]
	\centering
		\includegraphics[width=0.5\textwidth]{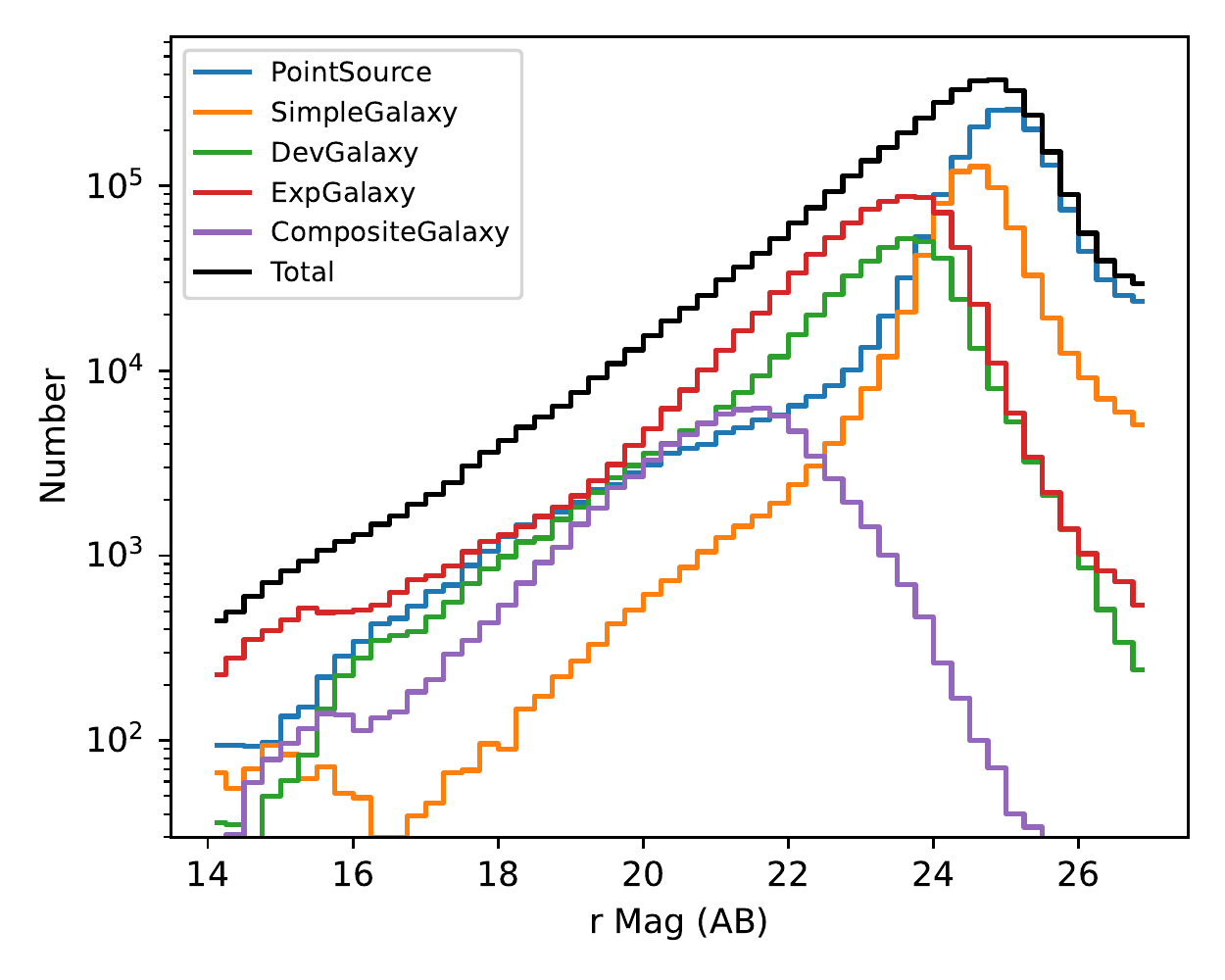}
		\caption{Distribution of source models as a function of HSC \R ~magnitude in T1. Sources are more resolved at brighter magnitudes. The unresolved PointSource model and barely resolved SimpleGalaxy model dominate fainter magnitudes of $r \gtrsim 24$. The resolved ExpGalaxy and DevGalaxy models dominate brighter magnitudes of $20 \lesssim r \lesssim 24$. The most complex CompositeGalaxy model is mostly found at $r \lesssim 22$.}
		\label{fig:mod}
\end{figure}

After creating the PSF models, source modeling is performed on a user-defined set of ``modeling bands'', which is usually a subset of all the available bands. In this catalog, our modeling bands include the DECam \R , \I , \Z ~and NEWFIRM \Ks ~bands. This is because all the detection bands have to be used in modeling so that every detected source can be modeled. Additionally, we include the HSC \R ~and \Z ~bands, which have improved depth and spatial resolution over the same DECam bands. The HSC \I ~band is not included since it does not fully cover the SHELA field. The modeling bands are then simultaneously fitted to determine the best-fit source profiles for each source. \farmer ~allows five different models to describe the source profiles:
\begin{enumerate}
    \item {\bf PointSource} models are used to describe unresolved sources. They are simply the PSF of each individual band, parametrized by the flux and source centroid.
    \item {\bf SimpleGalaxy} models are used to describe barely resolved sources. They are a special case of an exponential light profile, where the profile is circularly symmetric and has a fixed effective radius of $0\farcs45$.
    \item {\bf ExpGalaxy} models are exponential light profiles parametrized by the flux, effective radius, axis ratio, position angle and source centroid.
    \item {\bf DevGalaxy} models are de Vaucouleurs light profiles parametrized by  the flux, effective radius, axis ratio, position angle and source centroid.
    \item {\bf CompositeGalaxy} models are the concentric superposition of the ExpGalaxy and DevGalaxy models. They are parametrized by the source centroid, total flux and flux ratio between the two models, as well as the effective radius, axis ratio, position angle of each of the two models.
\end{enumerate}
The best fit model is chosen using a decision tree in \farmer . Briefly, it progresses from a simpler model to a more complex model if either the improvement in $\chi^2$ is greater than a user-defined threshold or the simpler model $\chi^2$ exceeds a certain value. We select the decision tree parameters by testing different parameters in steps on a small number of bricks. A set of parameters that balances between sufficient modeling and overfitting is chosen by visually inspecting the resulting source counts as a function of magnitude. The decision tree parameters are shown in Table \ref{tab:mod}. Figure \ref{fig:mod} shows the distribution of sources within a certain model as a function of HSC \R ~band magnitude. Faint sources with $r \gtrsim 24$ are mostly modelled by the PointSource and SimpleGalaxy models, as they are unresolved or barely resolved due to their low surface brightness. The ExpGalaxy and DevGalaxy models dominate an intermediate range of $20 \lesssim r \lesssim 24$, while the most complex CompositeGalaxy model are mostly found at $r \lesssim 22$, as the sources become better resolved at brighter magnitudes.

\begin{figure}[!tbp]
	\centering
		\includegraphics[width=0.5\textwidth]{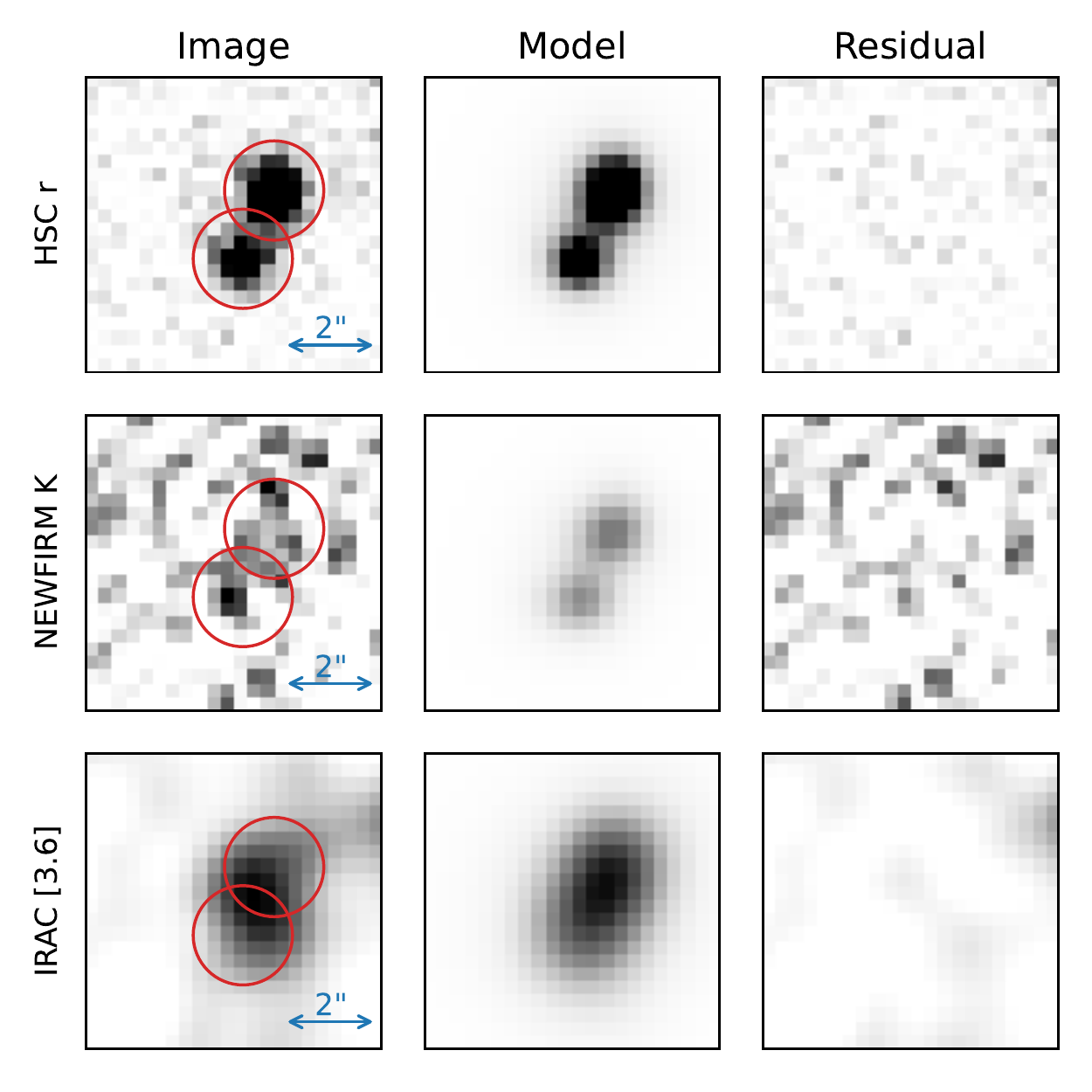}
		\caption{Demonstration of model fitting and forced photometry by \farmer . A blob containing a pair of neighboring sources separated by $\sim 1''$ is shown. The image, model and residual maps are shown from left to right. The HSC $r$, NEWFIRM \Ks ~and IRAC [3.6] bands are shown from top to bottom. $2''$-diamater apertures centered at the source positions are shown in red for reference. The sources are modeled simultaneously using the $rizK_s$ bands. Forced photometry is performed to all the bands by fixing the source profiles, convolved with the PSF of each band, and only allowing the flux normalizations to vary.}
		\label{fig:res}
\end{figure}

\begin{figure}[!tbp]
	\centering
		\includegraphics[width=0.5\textwidth]{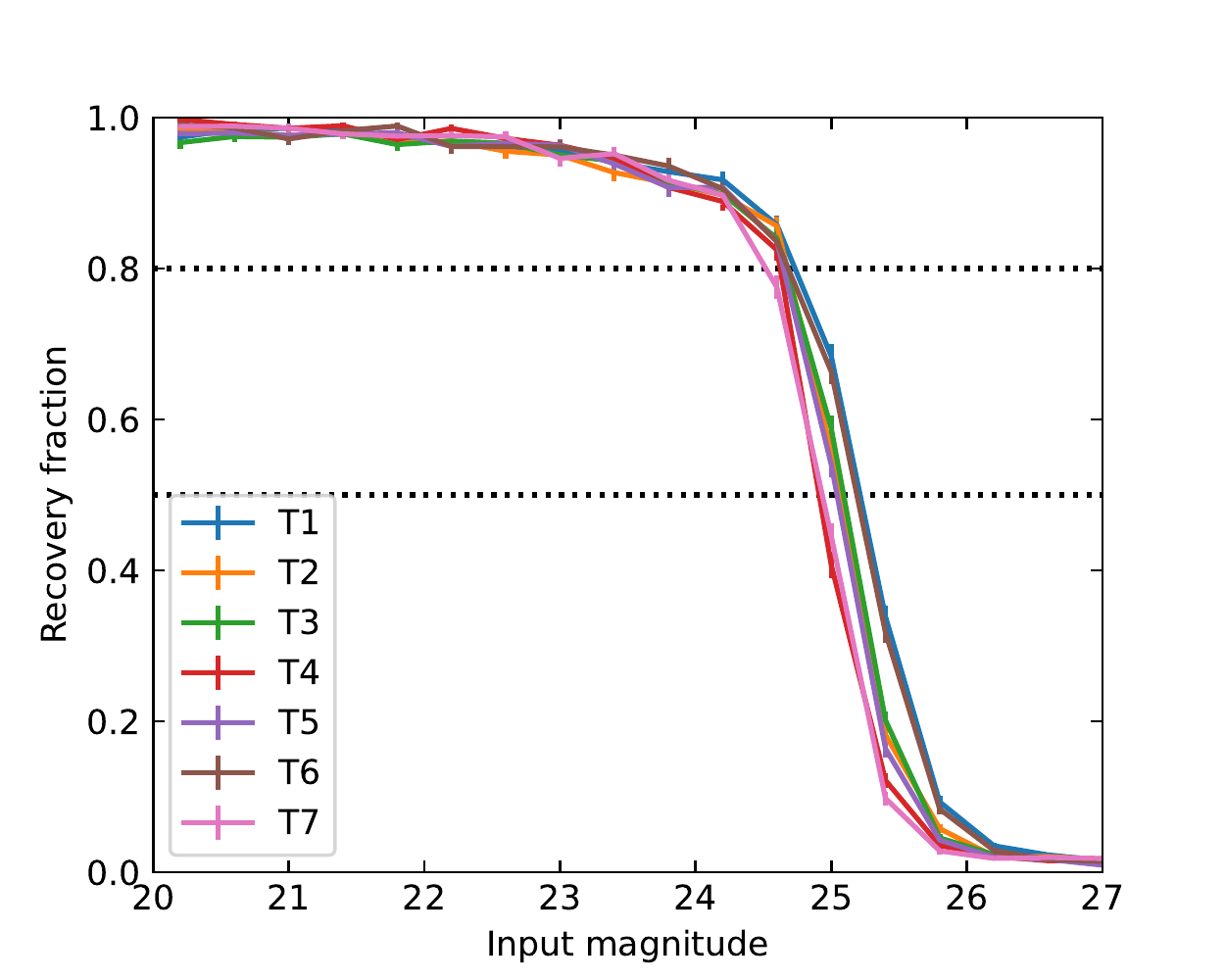}
		\caption{The fraction of simulated sources recovered as a function of input magnitude. The simulated sources have zero color, i.e. equal AB magnitudes between bands. The simulations show an $80\%$ ($50\%$) completeness threshold of $\sim 24.7$ ($\sim 25.1$) mag.}
		\label{fig:rec}
\end{figure}

\subsubsection{Forced photometry}

After a source profile model is chosen using the modeling bands, forced photometry is performed on all of the remaining bands that are not used in the modeling process. This is achieved by fitting each force photometric band individually by fixing the source profiles, convolved with the PSF of each band, and only allowing the flux normalizations to vary. Additionally, a small positional offset is allowed in each band to account for any potential imperfect astrometry. A Gaussian prior with a standard deviation of 0.3 pixels is applied to the position to prevent excessive offsets. In Figure \ref{fig:res}, we show the image, model map and residual map of an example blob containing a pair of sources separated by $\sim 1''$ for the HSC $r$, NEWFIRM $K_s$, and IRAC [3.6] bands. The forced photometry results in the complete photometry in all the photometric bands in this catalog. 

\subsubsection{Duplicates}
Since there is a small overlap between tiles, we identify and remove duplicate sources in these regions. Any source within $0\farcs5$ of a source in another tile is considered a duplicate, and the source that has a larger $\chi^2$ in the DECam \R-band is removed from the combined catalog. We note that the raw data covering these overlapping regions are identical. The only difference between the tile images is the different resampling schemes. We verified that the photometry of duplicate pairs are consistent within their uncertainties.

\subsection{Completeness}

We perform simulations to estimate the completeness of the catalog. We insert fake point sources with no color, i.e., equal AB magnitudes between bands, to the science images, and attempt to recover them using the same detection and photometry procedures as the catalog sources. For each tile, we randomly generate 15,000 locations and fluxes for source insertion. The fluxes are drawn from a power-law-plus-constant distribution between 20 and 27 mag. Point source profiles are constructed by multiplying the normalized position-dependent PSF stamps in each band with the desired flux. These fake point sources are then inserted into the science images of each band at the same set of locations. We then construct the \texttt{CHI-MEAN} detection image, generate bricks and detect sources in the same procedures. Figure \ref{fig:rec} shows for fraction of fake sources recovered as a function of magnitude. Our simulations show that our catalog is $80\%$ complete at $\approx 24.7$ mag. After this threshold, the completeness declines to $50\%$ at $\approx 25.1$ mag.

\subsection{Photometric Errors} \label{sec:err}

\begin{figure*}[!tbp]
	\centering
		\includegraphics[width=\textwidth]{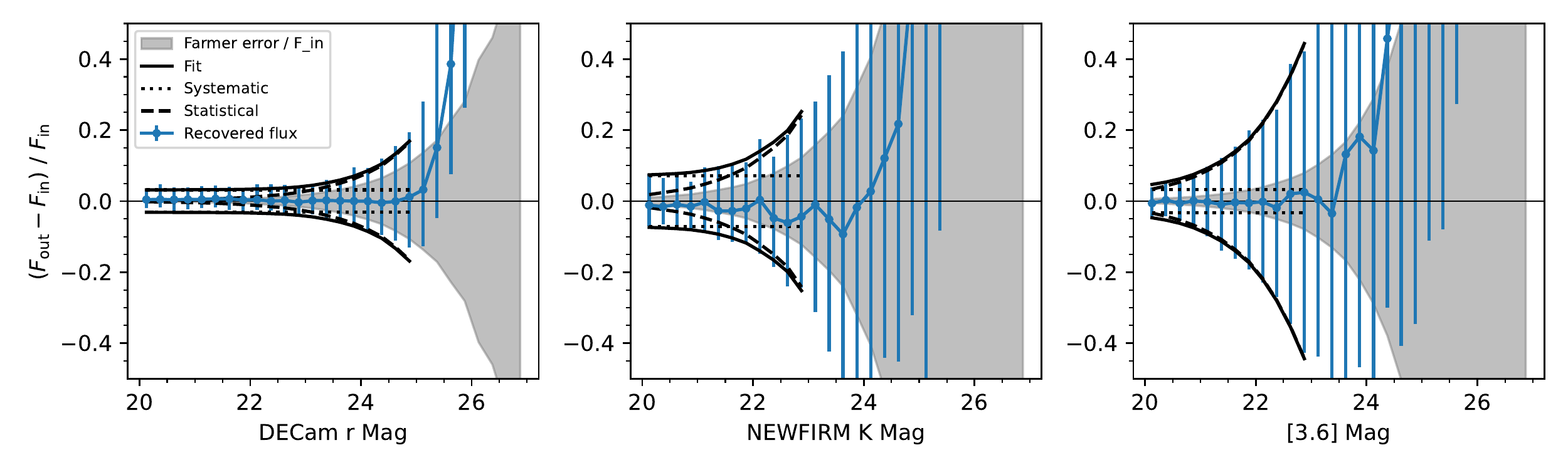}
		\caption{Demonstration of our measurement of flux error by simulation of injected sources in the DECam \R , NEWFIRM \Ks ~ and IRAC [3.6] bands in T1. The data points and error bars show the median and central 68-percentiles of the fractional flux difference between the injected and recovered sources, respectively, in each magnitude bin. The shaded region shows the median fractional error reported by \farmer ~($\sigma_\mathrm{Farmer} / F_\mathrm{in}$). The dotted, dashed and solid black lines show the best-fit systematic, statistical and total errors, respectively, as defined in Equation \ref{eq:fracerr}. The recovered error is substantially larger than the error reported by \farmer\ in the NEWFIRM and IRAC bands. This is likely due to a combination of larger PSF and native pixel sizes, resulting in more correlated pixels.}
		\label{fig:err}
\end{figure*}

We estimate point-source photometric errors using the same set of simulations and compare them with the flux errors reported by \farmer. After source detection, we proceed to measure the fluxes of the recovered sources in each band by performing our modeling and forced photometry on the blobs containing the recovered sources. We then measure the photometric errors by comparing the recovered and input fluxes in each tile and photometric band. As an example, Figure \ref{fig:err} shows the median and 68th-percentile of the fractional difference between the input and recovered fluxes 
as a function of the input magnitude for the DECam \R , NEWFIRM \Ks ~and IRAC [3.6] bands in T1. We also show the median error reported by \farmer ~in the same plot. The recovered error is generally larger than but follows a similar trend as the error reported by \farmer , showing that the errors reported by \farmer ~are underestimated. We therefore estimate the true photometric error by applying a correction to the error reported by \farmer . 

We parametrize the fractional photometric error as 
\begin{equation}
\label{eq:fracerr}
    \left(\frac{\sigma_{\mathrm{total}}}{f}\right)^2 = \sigma_\mathrm{sys}^2 + 
    C^2 \left(\frac{\sigma_{\mathrm{Farmer}}}{f}\right)^2,
\end{equation}
where $\sigma_\mathrm{tot}$ is the total photometric error, $f$ is the flux density, $\sigma_\mathrm{sys}$ is the dimensionless systematic error, $C$ is a dimensionless correction factor, and $\sigma_{\mathrm{Farmer}}$ is the error reported by \farmer ~in flux density. 
In each magnitude bin, we take half of the central 68-percentiles of the recovered fractional flux error as $\sigma_{\mathrm{total}}/f$, the median Farmer error as $\sigma_{\mathrm{Farmer}}$, and the flux corresponding to the magnitude bin center as $f$. We then perform a least squares fit to Equation \ref{eq:fracerr} to find the values of $\sigma_\mathrm{sys}$ and $C$. As the recovered fluxes fluctuate strongly at very faint magnitudes, we only make use of the data points brighter than than $m =$ 25 in our fitting for the optical bands and $m =$ 23 for the infrared bands.

The systematic error term is typically $\sim 5\%$ of the flux. A major source of systematic uncertainty is from the PSF model. \texttt{PSFEx} does not return uncertainties for the PSF stamps and the \texttt{Tractor} engine assumes the PSF models to be the truth. Therefore, the error reported by \farmer ~does not account for the effect of the uncertainty in the PSF model. 

The correction factor is typically $\sim 1.2-1.5$ for the optical and VICS82 bands. Larger correction factors of 2 and 5 are required for the NHS \Ks ~and IRAC bands, respectively. The need for a correction factor for the statistical error is likely due to correlated pixels. \farmer ~error, which is calculated by a weighted quadrature sum of pixel errors, gives the total error under the assumption of independent data points, i.e., uncorrelated pixel values, and therefore have zero covariance terms. The science images have been interpolated and resampled from their native pixel scales, which leads to correlation between neighbouring pixels. Furthermore, when the PSF is significantly wider than the pixel size, a considerable number of pixels within the PSF can become correlated. These factors invalidate the assumption of uncorrelated pixels, and lead to non-zero covariance terms and thus underestimation of the true statistical error. In fact, the bands that require the largest correction factors are the IRAC bands, followed by the NHS \Ks ~band, which have the largest PSF and native pixel size.

\begin{deluxetable*}{ccccccccccccccccccc}
\label{tab:depth}
\tablecaption{Point-source Detection Limits ($5\sigma$, AB)}
\tablehead{
\colhead{} & \multicolumn{6}{c}{DECam} & \multicolumn{5}{c}{HSC} & \multicolumn{2}{c}{VISTA} & \multicolumn{2}{c}{CFHT} & \colhead{NEWFIRM} & \multicolumn{2}{c}{IRAC} \\ 
\cmidrule(lr){2-7} \cmidrule(lr){8-12}  \cmidrule(lr){13-14} \cmidrule(lr){15-16} \cmidrule(lr){17-17} \cmidrule(lr){18-19}
\colhead{Tile} & \colhead{\U} & \colhead{\G} & \colhead{\R} & \colhead{\I} & \colhead{\Z} & \colhead{\Y} & \colhead{\G} & \colhead{\R} & \colhead{\I} & \colhead{\Z} & \colhead{\Y} & \colhead{$J$} & \colhead{$K_s$} & \colhead{$J$} & \colhead{$K_s$} & \colhead{$K_s$} & \colhead{[3.6]} & \colhead{[4.5]}
}
\startdata
\multicolumn{19}{c}{Simulation}  \\
\hline
1 & 25.1 & 25.5 & 25.1 & 24.8 & 24.3 & 22.9 & 26.1 & 25.6 & 25.1 & 24.8 & 23.4 & 22.7 & 21.9 & 22.4 & 21.8 & 22.6 & 21.9 & 21.9 \\
2 & 25.3 & 25.4 & 25.0 & 24.7 & 23.9 & 22.5 & 26.0 & 25.7 & 25.4 & 24.5 & 24.2 & 22.3 & 21.7 & - & - & 22.4 & 21.9 & 22.0 \\
3 & 25.4 & 25.4 & 25.1 & 24.7 & 24.1 & 22.3 & 26.1 & 25.8 & 25.7 & 24.5 & 24.3 & 22.3 & 21.8 & - & - & 22.2 & 21.9 & 22.0 \\
4 & 25.5 & 25.3 & 25.0 & 24.5 & 24.1 & 22.3 & 25.9 & 25.6 & 25.6 & 24.3 & 24.3 & 22.4 & 21.8 & 22.0 & 21.6 & 21.5 & 22.0 & 22.0 \\
5 & 25.4 & 25.4 & 25.0 & 24.5 & 24.1 & 22.1 & 25.7 & 25.2 & 25.0 & 24.6 & 24.1 & 21.7 & 21.7 & 22.0 & 21.7 & 21.9 & 22.0 & 22.0 \\
6 & 25.4 & 25.6 & 25.3 & 24.7 & 24.0 & 22.1 & 25.1 & 25.1 & 24.7 & 24.7 & 23.5 & 22.2 & 21.7 & - & - & 22.3 & 21.9 & 22.0 \\
7 & 25.3 & 25.1 & 24.8 & 24.4 & 24.1 & 22.3 & 25.4 & 24.9 & 24.8 & 24.6 & - & 22.1 & 21.0 & - & - & 22.5 & 21.9 & 22.0 \\
\hline
\multicolumn{19}{c}{\farmer}  \\
\hline
1 & 25.7 & 25.9 & 25.6 & 25.3 & 24.7 & 23.6 & 26.4 & 26.1 & 25.3 & 25.0 & 23.9 & 23.1 & 22.4 & 23.0 & 22.4 & 23.4 & 23.7 & 23.7 \\
2 & 25.7 & 25.8 & 25.4 & 25.0 & 24.5 & 22.5 & 26.3 & 26.2 & 25.7 & 24.8 & 24.5 & 22.7 & 22.1 & - & - & 23.2 & 23.7 & 23.7 \\
3 & 25.8 & 25.7 & 25.4 & 25.0 & 24.4 & 22.5 & 26.3 & 26.2 & 25.8 & 24.8 & 24.5 & 22.9 & 22.3 & - & - & 23.2 & 23.7 & 23.7 \\
4 & 25.7 & 25.7 & 25.2 & 24.8 & 24.4 & 22.4 & 26.3 & 26.2 & 25.8 & 24.7 & 24.5 & 22.9 & 22.3 & 22.6 & 22.1 & 23.2 & 23.7 & 23.7 \\
5 & 25.8 & 25.8 & 25.3 & 25.0 & 24.3 & 22.4 & 26.2 & 25.9 & 25.2 & 24.9 & 24.4 & 22.6 & 22.2 & 22.6 & 22.2 & 23.1 & 23.7 & 23.7 \\
6 & 25.8 & 26.2 & 25.6 & 25.0 & 24.5 & 22.4 & 26.2 & 25.6 & 25.1 & 25.0 & 23.8 & 22.8 & 22.2 & - & - & 23.1 & 23.7 & 23.7 \\
7 & 25.8 & 25.4 & 25.3 & 24.7 & 24.3 & 22.5 & 26.1 & 25.6 & 25.0 & 24.9 & - & 22.6 & 21.8 & - & - & 23.3 & 23.7 & 23.7 \\
\enddata
\tablecomments{Simulation detection limits are calculated using flux errors obtained by injecting and recovering simulated point sources in the science images. \farmer ~detection limits are calculated using the flux errors reported by \texttt{the Tractor}. See Section \ref{sec:err} for a detailed discussion of the differences between the two flux error measurements.}
\end{deluxetable*}

For each source, the total photometric error from the simulations is given by
\begin{equation}
\label{eq:error}
    \sigma_{\mathrm{total},i} = \sqrt{(\sigma_\mathrm{sys} f_i)^2 + 
    (C \sigma_{\mathrm{Farmer},i})^2},
\end{equation}
where $f_i$ and $\sigma_{\mathrm{Farmer},i}$ are the measured flux and error from \farmer ~for the source, and $\sigma_\mathrm{sys}$ and $C$ are the values measured by the simulations for the tile and band. In the catalog, we report the total error derived from Equation \ref{eq:error} as well as the error returned by \farmer. The total error is likely an conservative upper limit of the actual error, since a small fraction of recovered sources could be mismatched to a different source, which can lead to overestimation of the recovered flux error. We also note that we have only simulated point sources in our analysis, since there is an infinite number of possible resolved source profiles. As a result, these errors are only formally applicable to point sources.

\subsection{Detection Limits}

We estimate our point-source detection limits using the photometric errors obtained in the previous section. For each tile and band, we calculate the S/N by dividing the flux by the photometric error. We then estimate the $5\sigma$ detection limit by finding the median magnitude for point sources at $4.8 < S/N < 5.2$. We calculated detection limits both using the simulation-based errors and the values from \farmer , and the results are listed in Table \ref{tab:depth}. The simulation-based detection limits are generally shallower than \farmer -based ones by $\sim 0.3-0.5$ AB mag in the $ugrizY$-bands and VICS82 bands, $\sim 0.8$ AB mag in the NEWFIRM \Ks -band, and $\sim 1.7$ AB mag in both IRAC bands. 

Here we compare our simulation-based $5\sigma$ detection limits with previous studies. Comparing with the previous SHELA DECam catalog by \citet{wol19}, our DECam \U -band depths of $\sim 24.9-25.2$ AB mag are in agreement with their results, valued between their deeper ``sky aperture'' and shallower ``simulation'' detection limits. Our DECam $griz$-bands depths are generally $\sim 0.5$ AB mag deeper than the deeper ``sky aperture'' in \citet{wol19} thanks to our additional exposures. Our NEWFIRM \Ks -band reaches a depth of $\sim 22.3$ AB mag, similar to the 22.4 AB mag in \citet{ste21} based on $2''$-diameter apertures on the same \Ks ~dataset. Our IRAC 3.6 \um - and 4.5 \um -bands reach very uniform depths of $\sim 22.0$ AB mag, and are in excellent agreement with \citet{pap16} using the same IRAC dataset. In the VICS82 bands, our detection limits of $\sim 22.3$ and $\sim 21.7$ AB mag in $J$ and \Ks ~are $\sim 0.8$ AB mag deeper than those reported in \citet{vics82} based on $2''$-diameter apertures. This difference could be due to the better seeing in VICS82. Our model-based photometry and errors are based on fitting the light profile convolved with the PSF model, and are effectively an optimal extraction based on the actual seeing of the images. We speculate that the $2''$-diameter apertures in \citet{vics82} could be too large for their images, which have a typical seeing FWHM of $0.8-0.9''$ at VISTA, resulting in the inclusion of noise and thus a shallower measured detection limit. In comparison, the NEWFIRM \Ks ~images in \citet{ste21}, which are also measured by $2''$-diameter apertures and are in agreement with our results, have a typical seeing FWHM of $\sim 1.2''$.

\subsection{Number Counts}

\begin{figure}[!tbp]
	\centering
		\includegraphics[width=0.5\textwidth]{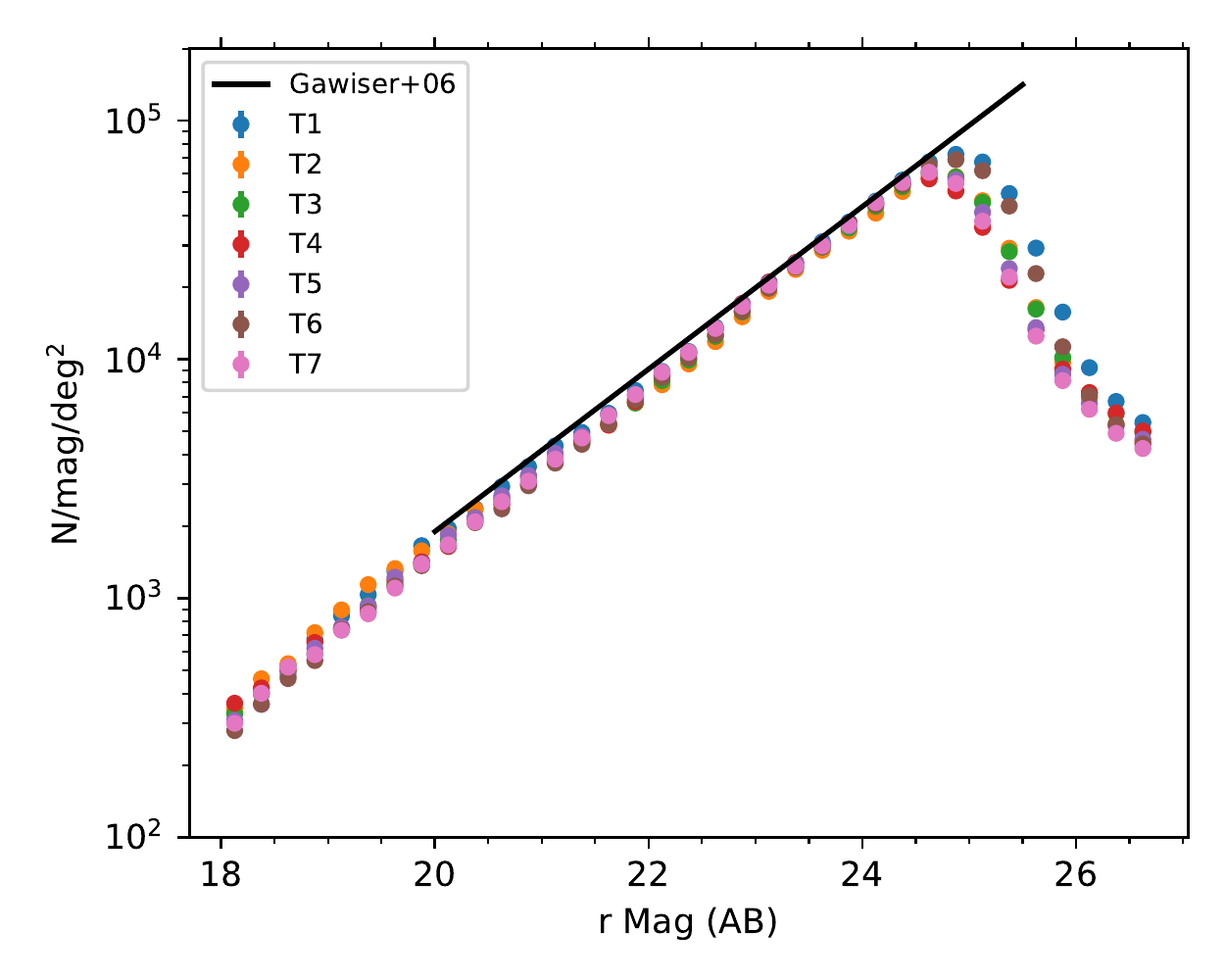}
		\caption{Differential number counts of galaxies versus DECam \R-band magnitude. Poisson error bars are smaller than the data point markers. The black line shows the best-fit line in \citet{gaw06}. Our measured number counts agree with the reported relation up to our $50\%$ completeness limit of $r \approx 25$, beyond which the number counts fall below the relation.}
		\label{fig:num}
\end{figure}

\begin{figure*}[!tbp]
	\centering
		\includegraphics[width=\textwidth]{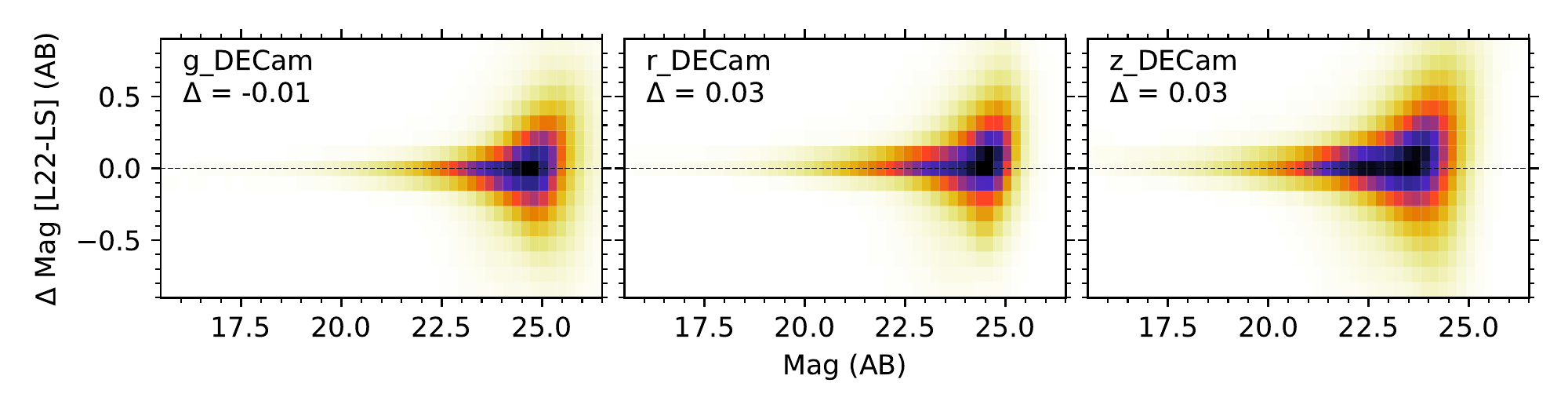}
		\caption{Difference between the measured magnitudes in this catalog (L22) and DECaLS DR9 (LS) as a function of magnitude in this catalog. The median magnitude difference is shown in each panel. DECaLS covers the $grz$-bands in DECam. Both catalogs employ model-based photometry. The photometry is in excellent agreement between the catalogs with a median offset of $\le 0.03$ mag.}
		\label{fig:ls}
\end{figure*}

An important test of the detection and photometry of the catalog is to compare the galaxy number counts to the literature. In Figure \ref{fig:num}, we plot the differential galaxy number counts against the measured DECam \R -band magnitude. We also plot in the same figure the best-fit line $\log(N) = -3.52 + 0.34 R$ to $20 < R < 24$ galaxies presented in \citet{gaw06}. Known stars in our catalog are removed by cross-matching with the SDSS DR17 star catalog. Our measured number counts agree with the reported relation up to $r \approx 25$, beyond which the number counts fall below the relation. The agreement at $r < 25$ shows that a smooth transition from the resolved to unresolved models is achieved during the modeling process. The decline beyond $r \approx 25$ is consistent with our completeness simulation results, which show an $80\%$ completeness threshold of $r \approx 24.7$. We use the DECam \R ~band here instead of the deeper HSC \R ~band because the HSC \R ~filter is bluer than the MOSAIC II $R$ filter in \citet{gaw06} by an effective wavelength of 330 \AA. The DECam filter, bluer by only $130$ \AA, provides a more direct comparison with \citet{gaw06}. The HSC number count follows a similar trend, but is slightly shifted towards fainter magnitudes.

\subsection{Comparison to DECaLS}

The DECaLS Survey \citep{decals} DR9 presents photometry using DECam imaging in the $grz$-bands observed through March 2019. The DECaLS data encompass most of the observing dates of those used in this catalog, and the footprint covers the SHELA field. Thus, it provides an independent reference catalog to verify the photometry in our catalog. Moreover, the DECaLS photometry is also based on source modeling by \texttt{the Tractor}, allowing an ``apples-to-apples'' comparison with this catalog. In Figure \ref{fig:ls}, we plot the difference in measured magnitude between this catalog and DECaLS as a function of the magnitude in the catalog for the DECam $grz$-bands. The photometry is in excellent agreement between the two catalogs. The median magnitude offset is $-0.01$ mag in \G , and 0.03 mag in \R ~and \Z .

\subsection{Comparison to Previous SHELA Catalog}

\begin{figure*}[!htbp]
	\centering
		\includegraphics[width=\textwidth]{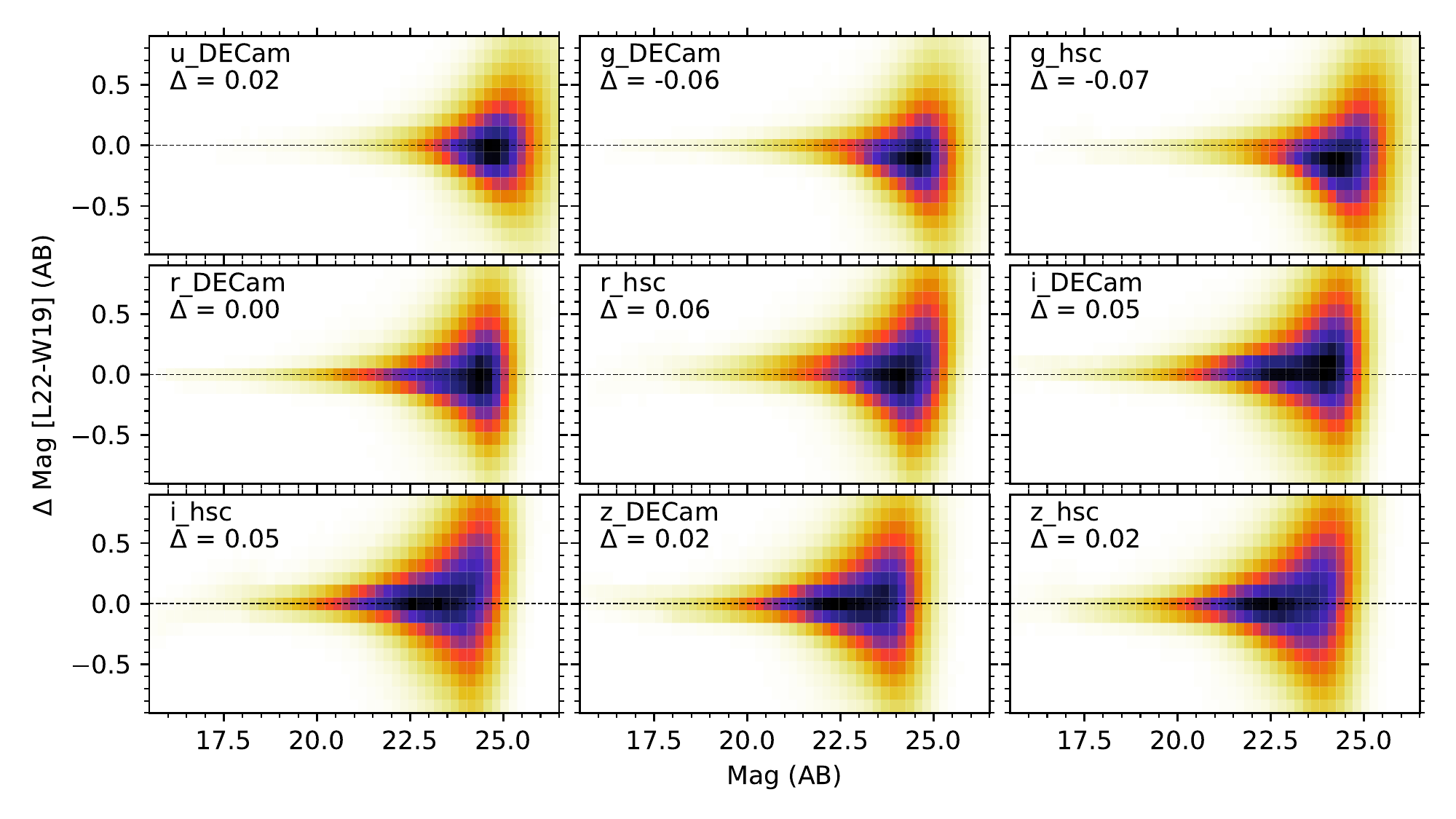}
		\caption{Comparison between the measured magnitudes in this catalog (L22) and the AUTO magnitudes using Kron apertures in \citet[][W19]{wol19}. The photometry in this catalog are generally within 0.1 mag of that in \citet{wol19}, with a absolute median offset of $< 0.07$ mag.} 
		\label{fig:wold}
\end{figure*}

We also compare our photometry with the previous aperture photometry catalog presented in the DECam catalog of \citet{wol19}. In Figure \ref{fig:wold}, we show the magnitude difference as a function of magnitude for the DECam $ugriz$ and HSC $griz$ bands. We use the ``AUTO'' magnitudes based on Kron apertures in \citet{wol19} for this comparison. Since this catalog includes additional photometric bands than \citet{wol19}, the HSC bands here are compared against the corresponding DECam bands. The magnitudes in this catalog are generally within 0.1 mag of those in \citet{wol19}, with a absolute median offset of $< 0.07$ mag.

\begin{figure}[!tbp]
	\centering
		\includegraphics[width=0.5\textwidth]{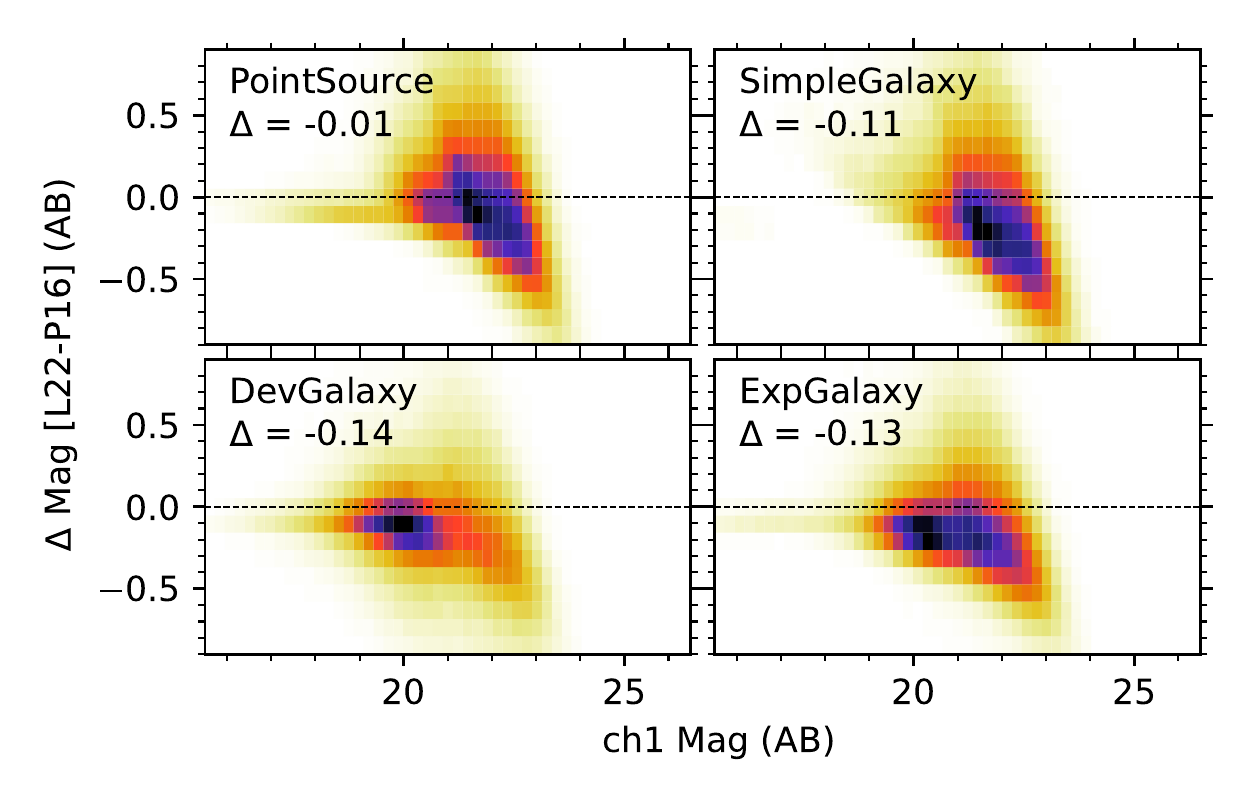}
		\caption{Comparison between IRAC [3.6] magnitudes in this catalog and the aperture-based SHELA IRAC catalog \citep[][P16]{pap16} for sources with different models. The magnitudes in P16 are measured in Kron apertures. The unresolved PointSource model results in fainter magnitudes, while the resolved SimpleGalaxy, DevGalaxy and ExpGalaxy models result in brighter magnitudes in this catalog. This suggests that model-based photometry is more capable at capturing all the flux from resolved source.}
		\label{fig:mod_mag}
\end{figure}

We further examine the relation between the magnitude offset and the light profile models of the sources in the IRAC bands, where source blending due to its lower resolution is expected to lead to the largest difference between model-based and aperture photometry. We compare the difference between the model-based magnitudes in this catalog and the ``AUTO'' magnitudes based on Kron apertures in the IRAC [3.6] band from the IRAC catalog of \citet{pap16}. In Figure \ref{fig:mod_mag}, we show the magnitude difference for sources with the PointSource, SimpleGalaxy, DevGalaxy and ExpGalaxy models. The PointSource model, which represents unresolved sources and dominates at $>24$ AB mag, have generally similar magnitudes as those in \citet{pap16}, showing a median offset of -0.01 mag. This suggests that point source fluxes are accurately measured in both model-based and aperture photometry. By contrast, the SimpleGalaxy, DevGalaxy and ExpGalaxy models, which represent resolved sources, have generally brighter magnitudes compared with \citet{pap16}, showing a median offset of $\sim -0.1$ mag. This suggests that a substantial fraction of the flux can be missed by the use of a fixed aperture on these resolved sources, and the modeling of the light profile and PSF is needed to capture all the flux from these sources. 

\section{Photometric Redshifts} \label{sec:photz}

\begin{figure*}[!th]
	\centering
		\includegraphics[width=\textwidth]{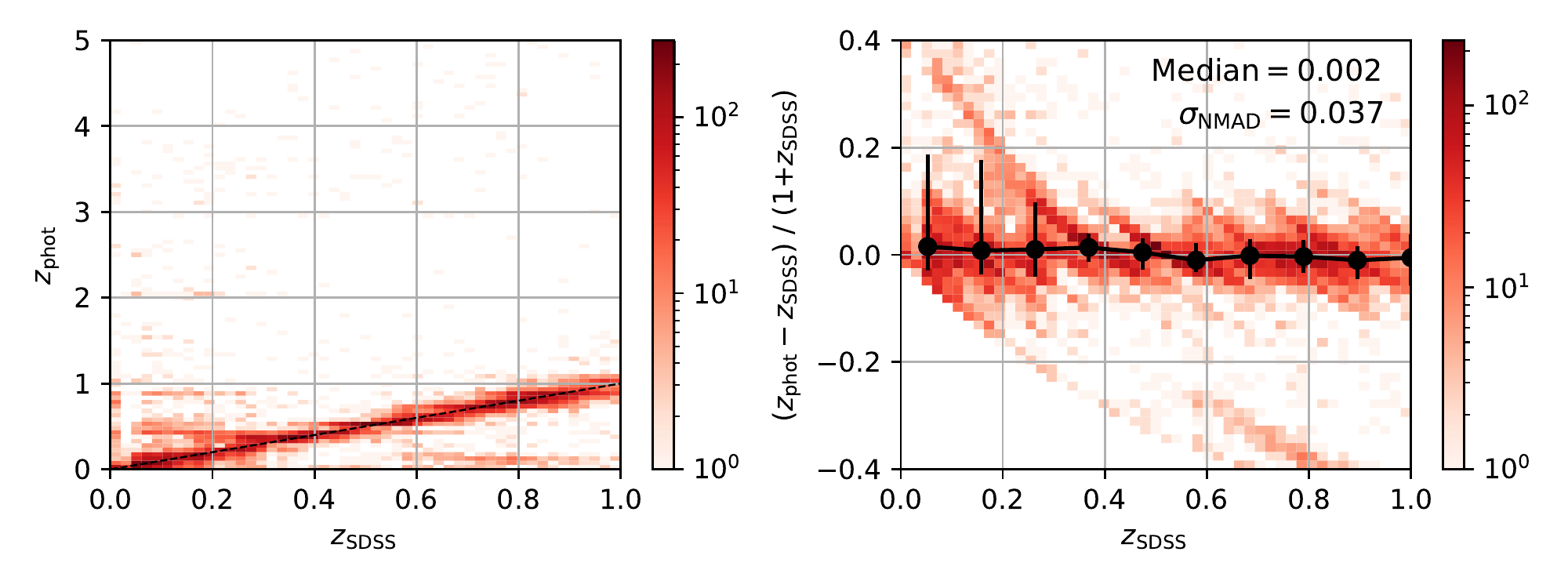}
		\caption{Comparison between photometric redshifts in this catalog and SDSS spectroscopic redshifts for 13,895 sources. In the left panel, the dashed black line shows where the photometric redshift and spectroscopic redshift are equal. In the right panel, the black points and error bars show the median and $68\%$-tiles, respectively. Outliers are mainly due to bright and highly resolved sources at $z \le 1$ that are not adequately modelled by the available light profiles and tend to get set to specific values of $z_\mathrm{phot}$.}
		\label{fig:specz}
\end{figure*}

We measure photometric redshifts using the photometric redshift code \texttt{EAZY} \citet{bra08}. We include fluxes in all the photometric bands that have $\chi^2 < 10$ and we use the photometric uncertainties inclusive of the systematic error terms and scaling factors derived from our simulations. We require the objects to have five valid flux measurements to ensure a well-cosntrained SED\null. We use a redshift grid from 0.01 to 15 with a step size of 0.01. We utilize the included EAZY template set, “tweak\_fsps\_QSF\_v12\_v3,” which uses a \citet{cha03} initial mass function, a \citet{kri13} dust attenuation law and solar metallicity. We also include an additional set of six templates in \citet{lar22} covering bluer colors than the included templates, which improves photometric redshifts for bluer galaxies.

We compare our photometric redshifts with spectroscopic redshifts in SDSS DR17. The vast majority of the sources with available spectroscopic redshifts are at $z \le 1$. In Figure \ref{fig:specz}, we compare the results for the 13,895 sources at $z \le 1$. The median $(z_\mathrm{phot} - z_\mathrm{SDSS})/(1+z_\mathrm{SDSS})$ is 0.002, and the normalized median absolute deviation (see \citealt{bra08}), defined as
\begin{equation}
    \sigma_\mathrm{NMAD} = 1.48 \times \mathrm{median} \left( \left| \frac{\Delta z - \mathrm{median}(\Delta z)}{1 + z_\mathrm{SDSS}} \right| \right),
\end{equation}
is 0.037. We find that $8 \%$ of the sources are $5\sigma_\mathrm{NMAD}$ outliers. The outliers are mostly attributed to imperfect source modeling for nearby galaxies, where bright, resolved structures are not adequately described by any of the light profiles used. An indicator of an inadequate fit is the $\chi^2$ of the source modeling process. We find that the median $\chi^2$ in the DECam \R-band for the $5\sigma_\mathrm{NMAD}$ outliers is 3.0 compared with 1.9 for all the sources, suggesting a generally worse fit in the outliers. This shows that the photometry of these outliers are less accurate than the rest of the sample. The number of outliers can be significantly reduced by applying a $\chi^2$ threshold during sample selection. We note that this effect is the most prominent for the bright, $z \le 1$ sources that are being compared here, since these sources can be sufficiently resolved that none of the model light profiles used can adequately describe the observed light distribution. In general, the $\chi^2$ values are close to unity at magnitudes $\gtrsim 22$ mag, indicating good photometry. 
Furthermore, the comparison here includes any available photometric redshift measurements in the catalog, while in actual applications, it is common to apply additional physically motivated selection criteria such as flux and/or S/N thresholds in specific photometric bands, which is expected to further reduce the outlier fraction. None of the $z_\mathrm{SDSS} \le 1$ galaxies compared here have a photometric redshift of $z_\mathrm{phot} \ge 10$.

\section{The SHELA Photometric Catalog}\label{sec:cat}

We publish with this paper the full SHELA photometric catalog. In Table \ref{tab:cat}, we show a sample of the SHELA photometric catalog. The catalog contains a unique source ID and J2000 coordinates for each source. We also report the best-fit light profile model selected by \farmer . For each band, we report the measured flux, the flux errors from both \farmer ~and our simulations, and the $\chi^2$ of the light profile fit. Fluxes and flux errors are in $\mu$Jy. We recommend using the simulation-based flux errors for most purposes. We advise against using fluxes with a corresponding $\chi^2 \gg 10$ for science. Finally, we list the photometric redshift information from EAZY, inculding the best-fit redshift, 68- and 95-percentiles, number of filters used in the fit, and the $\chi^2$ of the fit. Entries with no valid data are shown as $-99$.

\section{Summary}

In this paper, we have presented the $ugrizYJK_s$ plus 3.6 and 4.5 \um~ photometric catalog that reaches a $5\sigma$ depth $\sim 25.5$ AB mag for over four millions sources in the $\sim 27$ deg$^2$ SHELA field. We performed fully model-based photometry using \farmer , and validated our photometry with the model-based DECaLS DR9 catalog. We compared the model-based photometry with the previous aperture photometry catalogs. We find that model-based photometry can accurately measure point source fluxes and capture the full extended emission of resolved sources. We also presented photometric redshifts for the catalog, which show good agreement with available spectrosopic redshifts. The large area, multi-wavelength photometric catalog of SHELA will enable a wide range extragalactic studies.

\hfill 

We acknowledge that the location where most of this work took place, the University of Texas at Austin, sits on the Indigenous lands of Turtle Island, the ancestral name for what now is called North America. Moreover, we would like to acknowledge the Alabama-Coushatta, Caddo, Carrizo/Comecrudo, Coahuiltecan, Comanche, Kickapoo, Lipan Apache, Tonkawa and Ysleta Del Sur Pueblo, and all the American Indian and Indigenous Peoples and communities who have been or have become a part of these lands and territories in Texas. GCKL and SLF acknowledge support from the NSF through AAG grant awards 1908817 and 2009905. The Institute for Gravitation and the Cosmos is supported by the Eberly College of Science and the Office of the Senior Vice President for Research at the Pennsylvania State University.

This research draws upon DECam data as distributed by the Astro Data Archive at NSF's NOIRLab. NOIRLab is managed by the Association of Universities for Research in Astronomy (AURA) under a cooperative agreement with the National Science Foundation.

This project used data obtained with the Dark Energy Camera (DECam), which was constructed by the Dark Energy Survey (DES) collaboration. Funding for the DES Projects has been provided by the US Department of Energy, the US National Science Foundation, the Ministry of Science and Education of Spain, the Science and Technology Facilities Council of the United Kingdom, the Higher Education Funding Council for England, the National Center for Supercomputing Applications at the University of Illinois at Urbana-Champaign, the Kavli Institute for Cosmological Physics at the University of Chicago, Center for Cosmology and Astro-Particle Physics at the Ohio State University, the Mitchell Institute for Fundamental Physics and Astronomy at Texas A\&M University, Financiadora de Estudos e Projetos, Fundação Carlos Chagas Filho de Amparo à Pesquisa do Estado do Rio de Janeiro, Conselho Nacional de Desenvolvimento Científico e Tecnológico and the Ministério da Ciência, Tecnologia e Inovação, the Deutsche Forschungsgemeinschaft and the Collaborating Institutions in the Dark Energy Survey.

The Collaborating Institutions are Argonne National Laboratory, the University of California at Santa Cruz, the University of Cambridge, Centro de Investigaciones Enérgeticas, Medioambientales y Tecnológicas–Madrid, the University of Chicago, University College London, the DES-Brazil Consortium, the University of Edinburgh, the Eidgenössische Technische Hochschule (ETH) Zürich, Fermi National Accelerator Laboratory, the University of Illinois at Urbana-Champaign, the Institut de Ciències de l’Espai (IEEC/CSIC), the Institut de Física d’Altes Energies, Lawrence Berkeley National Laboratory, the Ludwig-Maximilians Universität München and the associated Excellence Cluster Universe, the University of Michigan, NSF’s NOIRLab, the University of Nottingham, the Ohio State University, the OzDES Membership Consortium, the University of Pennsylvania, the University of Portsmouth, SLAC National Accelerator Laboratory, Stanford University, the University of Sussex, and Texas A\&M University.

Based on observations at Cerro Tololo Inter-American Observatory, NSF’s NOIRLab (NOIRLab Prop. ID 2019B-0080; PI: C. Papovich), which is managed by the Association of Universities for Research in Astronomy (AURA) under a cooperative agreement with the National Science Foundation.

The Hyper Suprime-Cam (HSC) collaboration includes the astronomical communities of Japan and Taiwan, and Princeton University. The HSC instrumentation and software were developed by the National Astronomical Observatory of Japan (NAOJ), the Kavli Institute for the Physics and Mathematics of the Universe (Kavli IPMU), the University of Tokyo, the High Energy Accelerator Research Organization (KEK), the Academia Sinica Institute for Astronomy and Astrophysics in Taiwan (ASIAA), and Princeton University. Funding was contributed by the FIRST program from the Japanese Cabinet Office, the Ministry of Education, Culture, Sports, Science and Technology (MEXT), the Japan Society for the Promotion of Science (JSPS), Japan Science and Technology Agency (JST), the Toray Science Foundation, NAOJ, Kavli IPMU, KEK, ASIAA, and Princeton University. 

This paper makes use of software developed for Vera C. Rubin Observatory. We thank the Rubin Observatory for making their code available as free software at http://pipelines.lsst.io/.

This paper is based on data collected at the Subaru Telescope and retrieved from the HSC data archive system, which is operated by the Subaru Telescope and Astronomy Data Center (ADC) at NAOJ. Data analysis was in part carried out with the cooperation of Center for Computational Astrophysics (CfCA), NAOJ. We are honored and grateful for the opportunity of observing the Universe from Maunakea, which has the cultural, historical and natural significance in Hawaii. 

This work has made use of data from the European Space Agency (ESA) mission {\it Gaia} (\url{https://www.cosmos.esa.int/gaia}), processed by the {\it Gaia} Data Processing and Analysis Consortium (DPAC, \url{https://www.cosmos.esa.int/web/gaia/dpac/consortium}). Funding for the DPAC has been provided by national institutions, in particular the institutions participating in the {\it Gaia} Multilateral Agreement.

The Pan-STARRS1 Surveys (PS1) and the PS1 public science archive have been made possible through contributions by the Institute for Astronomy, the University of Hawaii, the Pan-STARRS Project Office, the Max-Planck Society and its participating institutes, the Max Planck Institute for Astronomy, Heidelberg and the Max Planck Institute for Extraterrestrial Physics, Garching, The Johns Hopkins University, Durham University, the University of Edinburgh, the Queen's University Belfast, the Harvard-Smithsonian Center for Astrophysics, the Las Cumbres Observatory Global Telescope Network Incorporated, the National Central University of Taiwan, the Space Telescope Science Institute, the National Aeronautics and Space Administration under Grant No. NNX08AR22G issued through the Planetary Science Division of the NASA Science Mission Directorate, the National Science Foundation Grant No. AST-1238877, the University of Maryland, Eotvos Lorand University (ELTE), the Los Alamos National Laboratory, and the Gordon and Betty Moore Foundation.





\addtocounter{table}{-1}
\movetabledown=5mm
\begin{longrotatetable} 
\begin{deluxetable*}{ccccccccccccc}
\label{tab:cat}
\tablecaption{SHELA Catalog Sample}
\tablehead{
\colhead{ID} & \colhead{R.A.} & \colhead{Decl.} & \colhead{Model} & \colhead{$f_\mathrm{u,DECam}$} & \colhead{$e^\mathrm{farmer}_\mathrm{u,DECam}$} & \colhead{$e^\mathrm{sim}_\mathrm{u,DECam}$} & \colhead{$\chi^2_\mathrm{u,DECam}$} & \colhead{$f_\mathrm{g,DECam}$} & \colhead{$e^\mathrm{farmer}_\mathrm{g,DECam}$} & \colhead{$e^\mathrm{sim}_\mathrm{g,DECam}$} & \colhead{$\chi^2_\mathrm{g,DECam}$} & \colhead{$f_\mathrm{r,DECam}$} \\
\colhead{(1)} & \colhead{(2)} & \colhead{(3)} & \colhead{(4)} & \colhead{(5)} & \colhead{(6)} & \colhead{(7)} & \colhead{(8)} & \colhead{(9)} & \colhead{(10)} & \colhead{(11)} & \colhead{(12)} & \colhead{(13)}
}
\startdata
100401830 & 15.933 & -0.526 & PointSource & 0.009 & 0.045 & 0.078 & 0.737 & -0.052 & 0.031 & 0.05 & 0.744 & 0.015 \\
100401831 & 15.882 & -0.526 & DevGalaxy & 0.038 & 0.044 & 0.076 & 1.117 & 0.119 & 0.03 & 0.048 & 0.896 & 0.816 \\
100401832 & 15.871 & -0.527 & PointSource & 0.321 & 0.038 & 0.07 & 0.961 & 0.288 & 0.028 & 0.045 & 1.009 & 0.414 \\
100401833 & 15.871 & -0.526 & ExpGalaxy & 0.84 & 0.082 & 0.156 & 1.135 & 0.863 & 0.054 & 0.09 & 0.996 & 1.371 \\
100401834 & 15.851 & -0.526 & PointSource & -0.081 & 0.041 & 0.071 & 1.065 & 0.148 & 0.025 & 0.041 & 1.002 & 0.524 \\
100401835 & 15.941 & -0.526 & PointSource & 0.138 & 0.034 & 0.061 & 1.148 & 0.311 & 0.024 & 0.04 & 1.45 & 0.86 \\
100401836 & 15.906 & -0.526 & ExpGalaxy & 1.321 & 0.044 & 0.127 & 0.829 & 1.736 & 0.026 & 0.059 & 0.965 & 3.904 \\
100401837 & 15.84 & -0.528 & SimpleGalaxy & 0.392 & 0.066 & 0.119 & 0.663 & 0.489 & 0.032 & 0.052 & 1.067 & 1.07 \\
100401838 & 15.84 & -0.528 & DevGalaxy & 3.561 & 0.059 & 0.292 & 0.968 & 4.179 & 0.027 & 0.108 & 0.906 & 6.082 \\
100401839 & 15.841 & -0.529 & PointSource & 0.368 & 0.05 & 0.092 & 0.814 & 0.674 & 0.025 & 0.043 & 0.943 & 1.515 \\
100401840 & 15.839 & -0.528 & SimpleGalaxy & 0.128 & 0.062 & 0.108 & 0.816 & 0.309 & 0.031 & 0.051 & 0.922 & 0.61 \\
100401841 & 15.838 & -0.527 & DevGalaxy & 2.725 & 0.064 & 0.238 & 0.831 & 10.376 & 0.038 & 0.253 & 1.023 & 44.144 \\
100401844 & 15.946 & -0.526 & PointSource & -0.056 & 0.037 & 0.065 & 0.753 & 0.1 & 0.028 & 0.045 & 0.259 & 0.219 \\
100401845 & 15.899 & -0.526 & PointSource & 0.257 & 0.039 & 0.071 & 1.025 & 0.282 & 0.025 & 0.042 & 0.497 & 0.415 \\
100401846 & 15.842 & -0.528 & SimpleGalaxy & 0.092 & 0.064 & 0.112 & 0.396 & 0.249 & 0.033 & 0.053 & 0.622 & 0.384 \\
100401847 & 15.842 & -0.527 & ExpGalaxy & 10.869 & 0.055 & 0.841 & 1.219 & 24.075 & 0.036 & 0.572 & 4.057 & 36.264 \\
100401848 & 15.841 & -0.526 & PointSource & 0.349 & 0.042 & 0.078 & 0.744 & 0.393 & 0.027 & 0.045 & 1.06 & 0.792 \\
100401850 & 15.919 & -0.526 & PointSource & 0.172 & 0.044 & 0.078 & 0.677 & 0.118 & 0.027 & 0.043 & 1.112 & 0.322 \\
100401851 & 15.891 & -0.526 & PointSource & 0.322 & 0.043 & 0.079 & 0.805 & 0.245 & 0.028 & 0.045 & 0.568 & 0.455 \\
100401852 & 15.846 & -0.526 & PointSource & 0.102 & 0.042 & 0.074 & 0.532 & 0.111 & 0.027 & 0.044 & 0.85 & 0.374
\enddata
\tablecomments{(1) Unique object ID number. (2) Object R.A. (J2000) in decimal degrees. (3) Object decl. (J2000) in decimal degrees. (4) Best-fit light profile model. (5 - 76) Measured flux in $\mu$Jy, flux error from \farmer ~in $\mu$Jy, flux error from our simulations in $\mu$Jy, and $\chi^2$ in the profile fitting for each band. (77) EAZY mimimum $\chi^2$ photometric redshift. (78 - 81) EAZY lower and upper 68- and 95-percentiles of photometric redshift. (82) Number of filters used in EAZY fit. (83) EAZY minimum $\chi^2$ value.}
\end{deluxetable*}
\end{longrotatetable}

\addtocounter{table}{-1}
\movetabledown=5mm
\begin{longrotatetable} 
\begin{deluxetable*}{ccccccccccccc}
\label{tab:cat}
\tablecaption{(Continued)}
\tablehead{
\colhead{$e^\mathrm{farmer}_\mathrm{r,DECam}$} & \colhead{$e^\mathrm{sim}_\mathrm{r,DECam}$} & \colhead{$\chi^2_\mathrm{r,DECam}$} & \colhead{$f_\mathrm{i,DECam}$} & \colhead{$e^\mathrm{farmer}_\mathrm{i,DECam}$} & \colhead{$e^\mathrm{sim}_\mathrm{i,DECam}$} & \colhead{$\chi^2_\mathrm{i,DECam}$} & \colhead{$f_\mathrm{z,DECam}$} & \colhead{$e^\mathrm{farmer}_\mathrm{z,DECam}$} & \colhead{$e^\mathrm{sim}_\mathrm{z,DECam}$} & \colhead{$\chi^2_\mathrm{z,DECam}$} & \colhead{$f_\mathrm{y,DECam}$} & \colhead{$e^\mathrm{farmer}_\mathrm{y,DECam}$} \\
\colhead{(14)} & \colhead{(15)} & \colhead{(16)} & \colhead{(17)} & \colhead{(18)} & \colhead{(19)} & \colhead{(20)} & \colhead{(21)} & \colhead{(22)} & \colhead{(23)} & \colhead{(24)} & \colhead{(25)} & \colhead{(26)}
}
\startdata
0.048 & 0.075 & 1.71 & 0.011 & 0.072 & 0.113 & 1.466 & -0.292 & 0.116 & 0.172 & 2.005 & 0.663 & 0.993 \\
0.048 & 0.08 & 1.183 & 2.26 & 0.081 & 0.142 & 0.889 & 5.478 & 0.13 & 0.378 & 1.016 & 6.553 & 0.827 \\
0.044 & 0.069 & 1.337 & 0.571 & 0.063 & 0.099 & 0.809 & 0.585 & 0.108 & 0.163 & 0.759 & -0.106 & 0.799 \\
0.094 & 0.154 & 1.209 & 2.46 & 0.136 & 0.223 & 0.962 & 3.154 & 0.24 & 0.401 & 0.925 & 5.08 & 1.65 \\
0.046 & 0.073 & 1.271 & 0.754 & 0.06 & 0.096 & 0.926 & 0.818 & 0.112 & 0.172 & 0.904 & 1.266 & 0.75 \\
0.039 & 0.067 & 1.077 & 2.241 & 0.064 & 0.119 & 0.931 & 4.924 & 0.099 & 0.327 & 1.123 & 6.765 & 0.75 \\
0.045 & 0.146 & 0.984 & 4.908 & 0.065 & 0.173 & 0.966 & 5.475 & 0.116 & 0.368 & 0.824 & 6.947 & 0.779 \\
0.058 & 0.097 & 0.781 & 1.805 & 0.082 & 0.137 & 0.941 & 3.212 & 0.149 & 0.291 & 0.748 & 4.264 & 0.961 \\
0.047 & 0.212 & 1.17 & 7.301 & 0.067 & 0.233 & 0.894 & 10.286 & 0.123 & 0.638 & 0.926 & 11.028 & 0.796 \\
0.047 & 0.089 & 0.835 & 1.667 & 0.06 & 0.105 & 1.1 & 1.569 & 0.111 & 0.189 & 0.983 & 3.031 & 0.782 \\
0.057 & 0.091 & 0.779 & 0.917 & 0.082 & 0.13 & 0.839 & 1.306 & 0.149 & 0.233 & 0.974 & 2.235 & 0.967 \\
0.065 & 1.447 & 1.13 & 74.321 & 0.094 & 2.133 & 1.183 & 99.153 & 0.175 & 5.897 & 1.056 & 112.973 & 1.101 \\
0.044 & 0.069 & 1.248 & 0.309 & 0.068 & 0.106 & 1.937 & 0.534 & 0.101 & 0.153 & 1.034 & -0.333 & 0.85 \\
0.042 & 0.067 & 0.463 & 0.685 & 0.063 & 0.1 & 0.587 & 1.227 & 0.11 & 0.178 & 1.19 & 1.358 & 0.748 \\
0.058 & 0.092 & 1.365 & 0.507 & 0.082 & 0.128 & 0.748 & 0.622 & 0.149 & 0.224 & 1.229 & 2.376 & 0.958 \\
0.059 & 1.189 & 5.45 & 46.151 & 0.083 & 1.327 & 4.818 & 48.856 & 0.151 & 2.911 & 2.691 & 53.52 & 0.946 \\
0.043 & 0.072 & 1.177 & 1.061 & 0.061 & 0.099 & 0.67 & 1.159 & 0.104 & 0.169 & 1.128 & 2.616 & 0.738 \\
0.042 & 0.066 & 1.281 & 0.283 & 0.062 & 0.097 & 1.379 & 0.301 & 0.1 & 0.148 & 1.02 & -0.526 & 0.795 \\
0.04 & 0.065 & 1.176 & 0.529 & 0.06 & 0.095 & 1.312 & 1.042 & 0.105 & 0.166 & 1.058 & 1.326 & 0.753 \\
0.044 & 0.071 & 1.333 & 0.359 & 0.064 & 0.101 & 0.843 & 0.427 & 0.108 & 0.162 & 0.742 & -0.214 & 0.793
\enddata
\end{deluxetable*}
\end{longrotatetable}

\addtocounter{table}{-1}
\movetabledown=5mm
\begin{longrotatetable} 
\begin{deluxetable*}{cccccccccccccccc}
\label{tab:cat}
\tablecaption{(Continued)}
\tablehead{
\colhead{$e^\mathrm{sim}_\mathrm{y,DECam}$} & \colhead{$\chi^2_\mathrm{y,DECam}$} & \colhead{$f_\mathrm{g,hsc}$} & \colhead{$e^\mathrm{farmer}_\mathrm{g,hsc}$} & \colhead{$e^\mathrm{sim}_\mathrm{g,hsc}$} & \colhead{$\chi^2_\mathrm{g,hsc}$} & \colhead{$f_\mathrm{r,hsc}$} & \colhead{$e^\mathrm{farmer}_\mathrm{r,hsc}$} & \colhead{$e^\mathrm{sim}_\mathrm{r,hsc}$} & \colhead{$\chi^2_\mathrm{r,hsc}$} & \colhead{$f_\mathrm{i,hsc}$} & \colhead{$e^\mathrm{farmer}_\mathrm{i,hsc}$} & \colhead{$e^\mathrm{sim}_\mathrm{i,hsc}$} & \colhead{$\chi^2_\mathrm{i,hsc}$} & \colhead{$f_\mathrm{z,hsc}$} & \colhead{$e^\mathrm{farmer}_\mathrm{z,hsc}$} \\
\colhead{(27)} & \colhead{(28)} & \colhead{(29)} & \colhead{(30)} & \colhead{(31)} & \colhead{(32)} & \colhead{(33)} & \colhead{(34)} & \colhead{(35)} & \colhead{(36)} & \colhead{(37)} & \colhead{(38)} & \colhead{(39)} & \colhead{(40)} & \colhead{(41)} & \colhead{(42)}
}
\startdata
1.809 & 2.87 & -0.019 & 0.032 & 0.043 & 2.764 & -0.048 & 0.035 & 0.056 & 1.342 & -0.002 & 0.047 & 0.058 & 0.543 & 0.018 & 0.081 \\
1.553 & 1.123 & 0.118 & 0.027 & 0.035 & 0.925 & 0.62 & 0.033 & 0.061 & 1.093 & 2.261 & 0.056 & 0.15 & 1.041 & 5.224 & 0.094 \\
1.455 & 0.88 & 0.267 & 0.022 & 0.03 & 1.39 & 0.297 & 0.027 & 0.045 & 1.346 & 0.273 & 0.041 & 0.053 & 1.692 & 0.469 & 0.071 \\
3.02 & 1.033 & 0.758 & 0.056 & 0.078 & 0.767 & 1.323 & 0.073 & 0.132 & 0.929 & 2.324 & 0.123 & 0.205 & 1.115 & 4.208 & 0.193 \\
1.368 & 1.003 & 0.138 & 0.021 & 0.029 & 1.388 & 0.429 & 0.027 & 0.048 & 1.254 & 0.495 & 0.042 & 0.06 & 0.846 & 0.697 & 0.076 \\
1.42 & 1.129 & 0.263 & 0.03 & 0.04 & 1.363 & 0.661 & 0.035 & 0.064 & 1.275 & 1.872 & 0.075 & 0.144 & 1.109 & 4.532 & 0.088 \\
1.475 & 0.967 & 1.617 & 0.027 & 0.061 & 1.029 & 3.664 & 0.036 & 0.179 & 0.943 & 4.623 & 0.054 & 0.28 & 1.018 & 5.309 & 0.086 \\
1.768 & 0.877 & 0.538 & 0.035 & 0.049 & 1.5 & 1.049 & 0.044 & 0.085 & 2.056 & 1.78 & 0.06 & 0.128 & 0.907 & 3.177 & 0.111 \\
1.582 & 0.974 & 4.039 & 0.031 & 0.13 & 0.999 & 5.682 & 0.037 & 0.27 & 1.489 & 6.726 & 0.046 & 0.4 & 1.435 & 9.674 & 0.088 \\
1.435 & 1.074 & 0.577 & 0.023 & 0.035 & 1.388 & 1.188 & 0.028 & 0.071 & 1.527 & 1.355 & 0.035 & 0.091 & 3.018 & 1.576 & 0.07 \\
1.765 & 1.157 & 0.303 & 0.034 & 0.046 & 1.169 & 0.667 & 0.043 & 0.076 & 1.02 & 0.947 & 0.058 & 0.091 & 1.051 & 0.898 & 0.114 \\
6.807 & 0.985 & 9.681 & 0.048 & 0.302 & 1.044 & 39.347 & 0.073 & 1.828 & 1.591 & 70.062 & 0.096 & 4.129 & 2.137 & 97.593 & 0.162 \\
1.548 & 1.316 & 0.038 & 0.024 & 0.031 & 1.544 & 0.199 & 0.026 & 0.043 & 0.238 & 0.189 & 0.073 & 0.091 & 1.669 & 0.464 & 0.091 \\
1.364 & 0.581 & 0.278 & 0.022 & 0.031 & 0.664 & 0.439 & 0.026 & 0.047 & 0.821 & 0.719 & 0.04 & 0.066 & 1.269 & 1.023 & 0.072 \\
1.751 & 0.739 & 0.149 & 0.035 & 0.046 & 1.07 & 0.341 & 0.043 & 0.071 & 0.939 & 0.442 & 0.058 & 0.077 & 1.476 & 0.589 & 0.117 \\
3.53 & 1.112 & 23.398 & 0.055 & 0.718 & 4.947 & 35.523 & 0.069 & 1.65 & 9.816 & 44.173 & 0.079 & 2.604 & 10.12 & 48.97 & 0.138 \\
1.352 & 0.946 & 0.302 & 0.022 & 0.031 & 1.292 & 0.567 & 0.028 & 0.052 & 0.971 & 0.841 & 0.036 & 0.066 & 1.094 & 0.954 & 0.072 \\
1.448 & 0.733 & 0.153 & 0.028 & 0.037 & 0.728 & 0.223 & 0.031 & 0.051 & 0.469 & 0.233 & 0.05 & 0.064 & 1.403 & 0.401 & 0.08 \\
1.373 & 1.086 & 0.273 & 0.023 & 0.031 & 1.095 & 0.368 & 0.027 & 0.046 & 0.72 & 0.647 & 0.041 & 0.063 & 1.543 & 0.747 & 0.07 \\
1.444 & 1.832 & 0.094 & 0.022 & 0.03 & 0.953 & 0.283 & 0.027 & 0.045 & 2.751 & 0.277 & 0.034 & 0.046 & 1.307 & 0.33 & 0.074
\enddata
\end{deluxetable*}
\end{longrotatetable}

\addtocounter{table}{-1}
\movetabledown=5mm
\begin{longrotatetable} 
\begin{deluxetable*}{cccccccccccccc}
\label{tab:cat}
\tablecaption{(Continued)}
\tablehead{
\colhead{$e^\mathrm{sim}_\mathrm{z,hsc}$} & \colhead{$\chi^2_\mathrm{z,hsc}$} & \colhead{$f_\mathrm{y,hsc}$} & \colhead{$e^\mathrm{farmer}_\mathrm{y,hsc}$} & \colhead{$e^\mathrm{sim}_\mathrm{y,hsc}$} & \colhead{$\chi^2_\mathrm{y,hsc}$} & \colhead{$f_\mathrm{K,NHS}$} & \colhead{$e^\mathrm{farmer}_\mathrm{K,NHS}$} & \colhead{$e^\mathrm{sim}_\mathrm{K,NHS}$} & \colhead{$\chi^2_\mathrm{K,NHS}$} & \colhead{$f_\mathrm{J,CFHT}$} & \colhead{$e^\mathrm{farmer}_\mathrm{J,CFHT}$} & \colhead{$e^\mathrm{sim}_\mathrm{J,CFHT}$} & \colhead{$\chi^2_\mathrm{J,CFHT}$} \\
\colhead{(43)} & \colhead{(44)} & \colhead{(45)} & \colhead{(46)} & \colhead{(47)} & \colhead{(48)} & \colhead{(49)} & \colhead{(50)} & \colhead{(51)} & \colhead{(52)} & \colhead{(53)} & \colhead{(54)} & \colhead{(55)} & \colhead{(56)}
}
\startdata
0.099 & 1.358 & 0.293 & 0.164 & 0.189 & 1.443 & -1.659 & 0.322 & 0.627 & 1.103 & -99.0 & -99.0 & -99.0 & -99.0 \\
0.3 & 1.118 & 6.788 & 0.259 & 1.046 & 0.995 & 25.276 & 0.365 & 1.946 & 0.993 & -99.0 & -99.0 & -99.0 & -99.0 \\
0.09 & 1.106 & 0.169 & 0.202 & 0.228 & 0.907 & -0.327 & 0.283 & 0.541 & 1.04 & -99.0 & -99.0 & -99.0 & -99.0 \\
0.325 & 0.937 & 4.381 & 0.559 & 0.902 & 1.168 & 14.56 & 0.57 & 1.51 & 1.008 & -99.0 & -99.0 & -99.0 & -99.0 \\
0.1 & 0.948 & 1.092 & 0.191 & 0.268 & 1.419 & 2.623 & 0.264 & 0.538 & 1.048 & -99.0 & -99.0 & -99.0 & -99.0 \\
0.263 & 1.229 & 5.358 & 0.199 & 0.824 & 1.01 & 13.799 & 0.257 & 1.107 & 1.361 & -99.0 & -99.0 & -99.0 & -99.0 \\
0.301 & 1.059 & 6.14 & 0.173 & 0.929 & 1.1 & 8.469 & 0.274 & 0.803 & 1.358 & -99.0 & -99.0 & -99.0 & -99.0 \\
0.216 & 1.004 & 4.722 & 0.258 & 0.757 & 1.285 & 17.147 & 0.344 & 1.397 & 1.505 & -99.0 & -99.0 & -99.0 & -99.0 \\
0.524 & 1.804 & 11.617 & 0.195 & 1.734 & 0.978 & 20.769 & 0.295 & 1.596 & 1.411 & -99.0 & -99.0 & -99.0 & -99.0 \\
0.12 & 1.372 & 2.079 & 0.155 & 0.353 & 0.985 & 2.791 & 0.274 & 0.561 & 1.126 & -99.0 & -99.0 & -99.0 & -99.0 \\
0.147 & 1.205 & 1.234 & 0.262 & 0.346 & 0.614 & 1.051 & 0.344 & 0.662 & 1.415 & -99.0 & -99.0 & -99.0 & -99.0 \\
5.18 & 1.563 & 121.248 & 0.323 & 17.956 & 1.127 & 306.261 & 0.392 & 22.023 & 6.388 & -99.0 & -99.0 & -99.0 & -99.0 \\
0.113 & 0.405 & 0.225 & 0.209 & 0.237 & 1.715 & 0.237 & 0.294 & 0.561 & 1.596 & -99.0 & -99.0 & -99.0 & -99.0 \\
0.103 & 1.371 & 1.631 & 0.151 & 0.295 & 0.995 & 1.78 & 0.294 & 0.575 & 0.804 & -99.0 & -99.0 & -99.0 & -99.0 \\
0.147 & 1.29 & -0.078 & 0.252 & 0.283 & 1.244 & 1.992 & 0.357 & 0.696 & 1.835 & -99.0 & -99.0 & -99.0 & -99.0 \\
2.603 & 3.894 & 55.671 & 0.286 & 8.249 & 1.75 & 64.695 & 0.352 & 4.698 & 1.492 & -99.0 & -99.0 & -99.0 & -99.0 \\
0.101 & 1.585 & 1.009 & 0.16 & 0.234 & 0.863 & 1.344 & 0.269 & 0.523 & 0.972 & -99.0 & -99.0 & -99.0 & -99.0 \\
0.1 & 0.715 & 0.452 & 0.156 & 0.187 & 1.178 & -0.289 & 0.277 & 0.53 & 1.45 & -99.0 & -99.0 & -99.0 & -99.0 \\
0.094 & 0.831 & 0.794 & 0.156 & 0.21 & 1.133 & 1.74 & 0.317 & 0.618 & 0.898 & -99.0 & -99.0 & -99.0 & -99.0 \\
0.092 & 1.172 & 0.262 & 0.156 & 0.18 & 1.028 & -0.0 & 0.283 & 0.54 & 0.597 & -99.0 & -99.0 & -99.0 & -99.0
\enddata
\end{deluxetable*}
\end{longrotatetable}

\addtocounter{table}{-1}
\movetabledown=5mm
\begin{longrotatetable} 
\begin{deluxetable*}{ccccccccccccc}
\label{tab:cat}
\tablecaption{(Continued)}
\tablehead{
\colhead{$f_\mathrm{K,CFHT}$} & \colhead{$e^\mathrm{farmer}_\mathrm{K,CFHT}$} & \colhead{$e^\mathrm{sim}_\mathrm{K,CFHT}$} & \colhead{$\chi^2_\mathrm{K,CFHT}$} & \colhead{$f_\mathrm{J,VISTA}$} & \colhead{$e^\mathrm{farmer}_\mathrm{J,VISTA}$} & \colhead{$e^\mathrm{sim}_\mathrm{J,VISTA}$} & \colhead{$\chi^2_\mathrm{J,VISTA}$} & \colhead{$f_\mathrm{K,VISTA}$} & \colhead{$e^\mathrm{farmer}_\mathrm{K,VISTA}$} & \colhead{$e^\mathrm{sim}_\mathrm{K,VISTA}$} & \colhead{$\chi^2_\mathrm{K,VISTA}$} & \colhead{$f_\mathrm{ch1}$} \\
\colhead{(57)} & \colhead{(58)} & \colhead{(59)} & \colhead{(60)} & \colhead{(61)} & \colhead{(62)} & \colhead{(63)} & \colhead{(64)} & \colhead{(65)} & \colhead{(66)} & \colhead{(67)} & \colhead{(68)} & \colhead{(69)}
}
\startdata
-99.0 & -99.0 & -99.0 & -99.0 & -0.222 & 0.411 & 0.627 & 0.498 & 1.668 & 0.755 & 1.109 & 0.428 & -0.729 \\
-99.0 & -99.0 & -99.0 & -99.0 & 9.907 & 0.534 & 0.886 & 0.69 & 22.912 & 0.956 & 1.698 & 0.896 & 33.791 \\
-99.0 & -99.0 & -99.0 & -99.0 & 1.285 & 0.386 & 0.589 & 0.489 & 0.206 & 0.699 & 1.025 & 1.009 & -2.765 \\
-99.0 & -99.0 & -99.0 & -99.0 & 10.092 & 1.155 & 1.796 & 0.787 & 12.987 & 2.104 & 3.132 & 0.887 & 22.713 \\
-99.0 & -99.0 & -99.0 & -99.0 & 0.461 & 0.393 & 0.599 & 0.756 & 0.795 & 0.757 & 1.11 & 0.592 & 1.971 \\
-99.0 & -99.0 & -99.0 & -99.0 & 6.716 & 0.406 & 0.663 & 0.955 & 12.336 & 0.766 & 1.236 & 0.645 & 13.013 \\
-99.0 & -99.0 & -99.0 & -99.0 & 5.596 & 0.441 & 0.7 & 0.956 & 5.09 & 0.81 & 1.206 & 1.222 & 5.948 \\
-99.0 & -99.0 & -99.0 & -99.0 & 5.965 & 0.719 & 1.115 & 0.788 & 18.291 & 1.291 & 2.041 & 1.333 & 31.584 \\
-99.0 & -99.0 & -99.0 & -99.0 & 12.053 & 0.563 & 0.958 & 0.949 & 16.343 & 0.971 & 1.579 & 1.035 & 31.866 \\
-99.0 & -99.0 & -99.0 & -99.0 & 1.197 & 0.457 & 0.698 & 0.762 & 1.086 & 0.788 & 1.156 & 0.618 & 4.211 \\
-99.0 & -99.0 & -99.0 & -99.0 & 2.354 & 0.762 & 1.164 & 0.284 & 0.106 & 1.212 & 1.776 & 1.101 & -2.117 \\
-99.0 & -99.0 & -99.0 & -99.0 & 160.535 & 0.799 & 5.794 & 0.837 & 291.758 & 1.441 & 12.384 & 1.0 & 164.433 \\
-99.0 & -99.0 & -99.0 & -99.0 & -0.507 & 0.408 & 0.622 & 0.714 & -1.282 & 0.824 & 1.208 & 0.242 & -0.249 \\
-99.0 & -99.0 & -99.0 & -99.0 & 0.358 & 0.376 & 0.572 & 0.563 & 2.35 & 0.685 & 1.009 & 1.232 & 4.021 \\
-99.0 & -99.0 & -99.0 & -99.0 & 2.096 & 0.72 & 1.099 & 0.788 & -0.362 & 1.365 & 2.001 & 0.625 & 2.889 \\
-99.0 & -99.0 & -99.0 & -99.0 & 57.239 & 0.756 & 2.326 & 1.08 & 58.244 & 1.337 & 3.126 & 0.764 & 39.279 \\
-99.0 & -99.0 & -99.0 & -99.0 & 0.82 & 0.433 & 0.661 & 0.586 & 0.741 & 0.712 & 1.044 & 0.493 & 0.726 \\
-99.0 & -99.0 & -99.0 & -99.0 & 0.817 & 0.356 & 0.543 & 1.308 & 0.972 & 0.656 & 0.962 & 0.913 & 0.003 \\
-99.0 & -99.0 & -99.0 & -99.0 & 1.894 & 0.379 & 0.581 & 0.713 & 0.082 & 0.69 & 1.012 & 0.822 & 0.851 \\
-99.0 & -99.0 & -99.0 & -99.0 & 0.602 & 0.398 & 0.607 & 1.418 & 1.637 & 0.81 & 1.19 & 0.9 & -0.01
\enddata
\end{deluxetable*}
\end{longrotatetable}

\addtocounter{table}{-1}
\movetabledown=5mm
\begin{longrotatetable} 
\begin{deluxetable*}{cccccccccccccc}
\label{tab:cat}
\tablecaption{(Continued)}
\tablehead{
\colhead{$e^\mathrm{farmer}_\mathrm{ch1}$} & \colhead{$e^\mathrm{sim}_\mathrm{ch1}$} & \colhead{$\chi^2_\mathrm{ch1}$} & \colhead{$f_\mathrm{ch2}$} & \colhead{$e^\mathrm{farmer}_\mathrm{ch2}$} & \colhead{$e^\mathrm{sim}_\mathrm{ch2}$} & \colhead{$\chi^2_\mathrm{ch2}$} & \colhead{$z_a$} & \colhead{$l_{68}$} & \colhead{$u_{68}$} & \colhead{$l_{95}$} & \colhead{$u_{95}$} & \colhead{$N_\mathrm{filt}$} & \colhead{$\chi^2_a$} \\
\colhead{(70)} & \colhead{(71)} & \colhead{(72)} & \colhead{(73)} & \colhead{(74)} & \colhead{(75)} & \colhead{(76)} & \colhead{(77)} & \colhead{(78)} & \colhead{(79)} & \colhead{(80)} & \colhead{(81)} & \colhead{(82)} & \colhead{(83)}
}
\startdata
0.322 & 1.766 & 0.072 & -1.026 & 0.302 & 1.674 & 0.031 & 6.62 & 3.441 & 12.373 & 0.329 & 14.624 & 16.0 & 18.686 \\
0.186 & 1.54 & 0.41 & 26.067 & 0.238 & 1.527 & 0.257 & 0.98 & 0.933 & 1.085 & 0.877 & 1.159 & 16.0 & 4.349 \\
0.197 & 1.081 & 0.224 & -2.458 & 0.209 & 1.158 & 0.823 & 0.71 & 0.184 & 0.86 & 0.029 & 1.359 & 16.0 & 25.704 \\
0.317 & 1.903 & 0.658 & 25.183 & 0.346 & 2.056 & 2.46 & 1.46 & 1.052 & 1.473 & 0.807 & 1.696 & 16.0 & 9.656 \\
0.223 & 1.222 & 0.273 & 2.918 & 0.222 & 1.234 & 0.307 & 3.76 & 3.406 & 4.049 & 0.35 & 4.257 & 16.0 & 12.265 \\
0.226 & 1.317 & 0.352 & 17.91 & 0.224 & 1.352 & 0.959 & 1.12 & 1.063 & 1.155 & 1.012 & 1.198 & 16.0 & 19.752 \\
0.18 & 1.008 & 0.431 & 5.108 & 0.212 & 1.185 & 1.294 & 0.5 & 0.439 & 0.571 & 0.37 & 0.631 & 16.0 & 6.428 \\
0.22 & 1.616 & 0.699 & 32.944 & 0.201 & 1.48 & 2.493 & 1.26 & 1.175 & 1.383 & 1.07 & 1.468 & 16.0 & 2.585 \\
0.196 & 1.529 & 3.224 & 34.995 & 0.184 & 1.452 & 5.194 & 1.28 & 1.159 & 1.286 & 1.095 & 1.373 & 16.0 & 18.946 \\
0.194 & 1.075 & 1.153 & 10.996 & 0.182 & 1.062 & 6.152 & 3.81 & 3.727 & 4.008 & 3.603 & 4.119 & 16.0 & 28.584 \\
0.22 & 1.206 & 0.395 & -1.335 & 0.196 & 1.089 & 0.959 & 0.01 & 0.066 & 0.528 & 0.016 & 0.72 & 16.0 & 12.114 \\
0.24 & 5.769 & 1.474 & 138.974 & 0.213 & 4.276 & 2.637 & 0.37 & 0.309 & 0.417 & 0.251 & 0.462 & 16.0 & 1.449 \\
0.297 & 1.628 & 0.08 & 0.55 & 0.282 & 1.565 & 0.152 & 4.22 & 0.371 & 4.427 & 0.07 & 4.723 & 16.0 & 11.391 \\
0.25 & 1.379 & 0.142 & 1.575 & 0.256 & 1.422 & 0.383 & 1.05 & 0.865 & 1.194 & 0.129 & 1.333 & 16.0 & 7.751 \\
0.231 & 1.27 & 1.307 & 1.243 & 0.216 & 1.196 & 0.067 & 0.71 & 0.527 & 3.691 & 0.145 & 4.134 & 16.0 & 11.971 \\
0.215 & 1.784 & 0.801 & 28.336 & 0.201 & 1.392 & 1.231 & 0.11 & 0.061 & 0.18 & 0.018 & 0.237 & 15.0 & 1.808 \\
0.2 & 1.093 & 0.811 & 1.262 & 0.188 & 1.041 & 0.2 & 0.42 & 0.398 & 0.66 & 0.169 & 0.745 & 16.0 & 10.175 \\
0.267 & 1.465 & 0.289 & -0.167 & 0.237 & 1.311 & 1.212 & 0.39 & 0.235 & 1.291 & 0.035 & 3.769 & 16.0 & 5.789 \\
0.253 & 1.387 & 0.142 & 2.573 & 0.235 & 1.304 & 0.441 & 0.88 & 0.254 & 1.078 & 0.032 & 1.432 & 16.0 & 9.086 \\
0.267 & 1.461 & 0.203 & 0.719 & 0.23 & 1.274 & 0.2 & 0.45 & 0.36 & 4.002 & 0.145 & 4.277 & 16.0 & 5.146
\enddata
\end{deluxetable*}
\end{longrotatetable}

\bibliography{mybib}{}
\bibliographystyle{aasjournal}

\end{document}